\documentclass[a4paper,11pt]{article}
\usepackage{jcappub} % for details on the use of the package, please see the JINST-author-manual
\usepackage{aas_macros} 
\usepackage{xspace}

\arxivnumber{2309.05618} % Only if you have one
\title{Progress in Direct Measurements of the Hubble Constant}

\newcommand{\cchp}{\,CCHP\xspace}                        
                       
\newcommand{\gaia}{{\it Gaia}\xspace}            
\newcommand{\ho}{$H_{0}$\xspace}                  
\newcommand{\hounits}{\,km\,s$^{-1}$\,Mpc$^{-1}$\xspace}   
\newcommand{\hst}{{HST}\xspace}              
\newcommand{\hstwfciii}{{\it HST/WFC3}\xspace}    
\newcommand{\hstacs}{{\it HST/ACS}\xspace} 
\newcommand{\jagb}{{\it JAGB}\xspace}  
\newcommand{\jwst}{{\it JWST}\xspace}

\newcommand{\nircam}{{\it NIRCam}\xspace}  
\newcommand{\niriss}{{\it NIRISS}\xspace}

\newcommand{\sne}{SNe~Ia\xspace}                   
\newcommand{\sn}{SN~Ia\xspace}                   
               
\newcommand{\shoes}{{\it SHoES}\xspace}            
\newcommand{\ngc}{NGC\,\xspace}

\author[a]{Wendy L. Freedman}
\affiliation[a]{The Department of Astronomy \& Astrophysics, and the  Kavli Institute for Cosmological Physics, University of Chicago, 5640 S. Ellis Ave., Chicago, IL, 60637}
\author[b]{and Barry F. Madore}
\affiliation[b]{The Observatories, Carnegie Institution for Science, 813 Santa Barbara St., Pasadena, CA, 91101, and the Department of Astronomy \& Astrophysics,  University of Chicago, 5640 S. Ellis Ave., Chicago, IL, 60637}
\emailAdd{wfreedman@uchicago.edu}

\abstract{One of the most exciting and pressing issues in cosmology today is the discrepancy between some measurements of the local Hubble constant and other values of the expansion rate  inferred from the observed temperature and polarization fluctuations in the cosmic microwave background (CMB) radiation. Resolving these differences holds the potential for the discovery of new physics beyond the standard model of cosmology: Lambda Cold Dark Matter ($\Lambda$CDM), a successful model that has been in place for more than 20 years. Given both the fundamental significance of this outstanding discrepancy, and the many-decades-long effort to increase the accuracy of the extragalactic distance scale, it is critical to demonstrate that the local measurements are convincingly free from residual systematic errors. We review the progress over the past quarter century in measurements of the local value of the Hubble constant, and discuss remaining challenges. Particularly exciting are  new data  from the James Webb Space Telescope (\jwst), for which we present an overview of our program and first results. We focus in particular on Cepheids and the Tip of the Red Giant Branch (TRGB) stars, as well as a relatively new method, the JAGB (J-Region Asymptotic Giant Branch) method, all  methods that currently exhibit the demonstrably smallest statistical and systematic uncertainties. \jwst is  delivering high-resolution near-infrared imaging data to both test for and to address directly several of the systematic uncertainties that have historically limited the accuracy of extragalactic distance scale measurements (e.g., the dimming effects  of interstellar dust, chemical composition differences in the atmospheres of stars, and the crowding and blending of Cepheids contaminated by nearby previously unresolved stars). For the first galaxy in our program, \ngc 7250, the high-resolution \jwst images demonstrate that many of the Cepheids observed with the Hubble Space Telescope (\hst) are significantly crowded by nearby neighbors.  Avoiding the more significantly crowded variables, the scatter in the \jwst near-infrared (NIR) Cepheid PL relation is decreased by a factor of two compared to those from \hst, illustrating the power of \jwst for improvements to local measurements of \ho. Ultimately, these data will  either confirm the standard model, or provide robust evidence for the inclusion of additional new physics.   }
\begin{document}
\maketitle
\flushbottom

\section{Introduction}
\label{sec:intro}

The year 2023 marks 100 years since Edwin Hubble's famous discovery of a single Cepheid variable in the Andromeda galaxy. Hubble's subsequent measurements of extragalactic distances were based (in part) on the Cepheid Period-Luminosity (PL) relation, aka the Leavitt Law \cite{leavitt_1908}. Correlating these distances with spectral measurements of radial (line-of-sight) velocities \cite{slipher_1915}, 
ultimately led to the discovery of the expansion of the universe in 1929 \cite{hubble_1929}, and ushered in modern cosmology.\footnote{It is now appreciated that Lemaitre \cite{lemaitre_1927}  had earlier found a mathematical solution for an expanding universe, recognizing that it provided a natural explanation for the observed recession velocities of galaxies, but these results were published in French in the Annals of the Scientific Society of Brussels, and at that time were not widely accessible.}

At the time of its launch in 1990, one of the highest priorities for the Hubble Space Telescope (\hst) was to convincingly measure the current rate of the expansion of the universe, the Hubble constant (\ho), to an accuracy of 10\%. In a Cepheid-based calibration, the Hubble Key Project team in 2001 obtained a value of \ho = 72 $\pm$ 3 (statistical) $\pm$ 7 (systematic) \cite{freedman_2001}. Two additional decades of effort with \hst, $Spitzer$, and many additional ground-based telescopes,  subsequently improved the measurements of \ho, with estimated accuracies currently falling in the 2-5\% range \cite{divalentino_2021}. The Cepheid calibration of \ho \cite{freedman_2012, riess_2022} continues to yield values of \ho $\sim$ 73 \hounits, whereas measurements using the tip of the red giant branch (TRGB) \cite{freedman_2020, dhawan_2023} yield slightly lower values, closer to 70 \hounits.
\begin{figure}
    \centering
\includegraphics[width=\columnwidth]{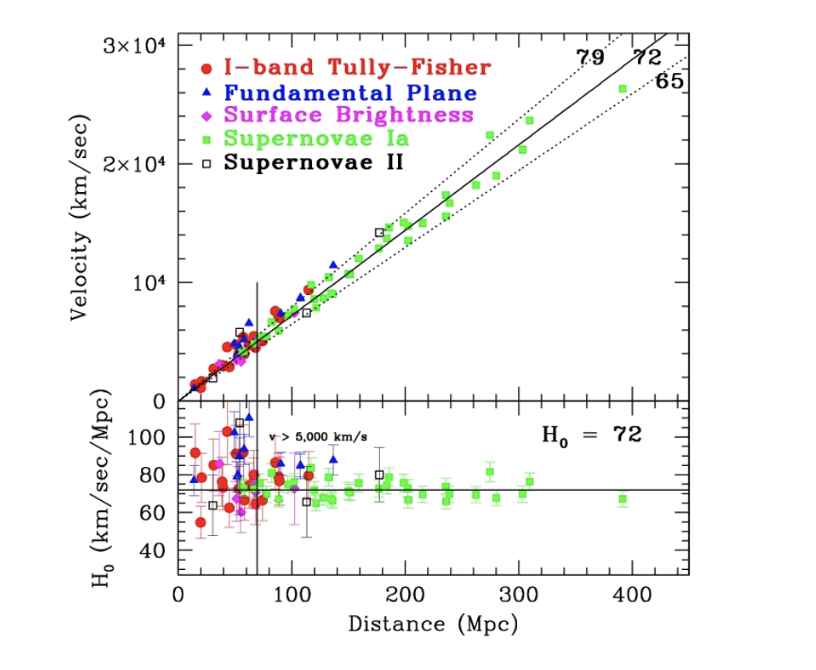}
    \caption{The Key Project Hubble Diagram: Distances in Mpc vs flow-corrected velocities in km/s. Secondary distance indicators used in this diagram are color-coded and identified in the legend at the top left of the figure. Residuals from a fit of \ho = 72 $\pm$ 7 km/s/Mpc are shown in the lower panel. Adapted from \cite{freedman_2001}. SNe extend out the farthest of the secondary methods and have the lowest dispersion. The other methods have a greater scatter owing to peculiar velocities and bulk flows. The vertical line denotes a velocity of 5000 km/s.} 
\label{fig:KP1}
\end{figure}
Recent estimates of \ho from CMB measurements have extremely high precision, with results from the Planck satellite \cite{planck_2018} yielding \ho = 67.4 $\pm$ 0.5 \hounits (better than 1\%)\footnote{For a discussion of possible systematics in the CMB analysis see \cite{yeung_chu_2022} and references therein.}.This level of precision is new for the field of observational cosmology, where until as recently as a couple of decades ago, a factor-of-two uncertainty had persisted for several decades. In a sense, this new level of precision has led to high expectations for other types of cosmological measurements. Yet, obtaining equally high precision observations of the local measurements of \ho remains a formidable challenge.  At face value, the inconsistency between the local value of the Hubble constant and the cosmologically modeled value could be interpreted as an inadequacy in the theory, thereby begging the questions: Is cosmology in a crisis? And is our current model of the universe now in need of new physics?  

While acoustic oscillations of the ionized plasma in the early universe are well understood and based on linear physics, it is important to keep in mind that the  astrophysics of stellar distance indicators is less predictive from first principles; and the requirement of accurate {\it absolute} calibrations of the local distance scale at a comparable (1\%) level, with the identification and elimination of systematic effects for evolving stars (which may be located in dusty, crowded regions) are tall orders. Given the current challenges in obtaining percent level accuracy in the local distance measurements, it may be premature to be claiming either confirmation, or the refutation, of the need for  physics beyond the standard model \cite{freedman_2017}. These remaining challenges underscore the need for a definitive measure of \ho locally, which in turn demands a complete and independently confirmed assessment of its total (statistical and systematic) uncertainties \cite{lopez-corredoira_2022}. 

Ascertaining whether additional physics is required beyond the standard $\Lambda$CDM model, although a challenge, is a surmountable observational issue. It will require the establishment of several independent calibrations of the local distance scale, each with high precision,  to provide robust constraints on the overall systematics. 

In this review, we briefly summarize historical efforts in the direct, local (astrophysical) measurements of \ho, present highlights from the past two decades in refining the measurements, discuss the substantial progress in overcoming systematic uncertainties, and note where we can expect most progress in the coming years, including very exciting new results from \jwst. In an appendix, we compare and contrast the strengths and weaknesses of the most promising methods in use today for measuring distances in the local universe. The prospects are good for a resolution to the {\it local} (distance scale) version of the \ho tension. The past 20 years have been referred to as the era of `precision cosmology'. We must now ensure that we have convincingly entered the era of `accurate cosmology'.

\section{The Landscape at the Turn of the Century: The Hubble Key Project}

The launch of \hst in 1990 provided the opportunity to undertake a major program to calibrate the extragalactic distance scale. The \hst Key Project was designed to measure the Hubble constant to a total (statistical plus systematic) uncertainty of $\pm$10\% \cite{freedman_2001}. Given that the dominant sources of error were clearly systematic in nature, the approach taken in the Key Project was to measure \ho by intercomparing several different methods, each having minimally overlapping systematics. The goal was to extend and apply the Cepheid distance scale beyond what could be achieved from the ground, and then to assess and to quantify the overall systematic errors in the measurement of \ho. Observations were obtained in the V band (F555W; 12 epochs within a 60-day window + 1 additional epoch a year later, to avoid aliasing effects) and the I band (F814W; 4 epochs). The roll angle of the telescope was held fixed for all of the observations to maximize overlap of the different epochs and to facilitate the photometric measurements. Data were taken with a power-law spacing to minimize aliasing effects \cite{madore_freedman_2005}. In addition, a test for the metallicity dependence of the Cepheid PL relation was undertaken.

Cepheids are supergiants, but they are still not sufficiently bright that they can be used to determined distances far enough away to sample the unperturbed cosmic Hubble flow.  Large-scale flows generated by major clusters, filaments and voids induce so-called ``peculiar velocities'' on one another and on individual field galaxies.  This ubiquitous source of noise in the velocity field must either be modelled out, averaged over large samples, or diminished in its relative impact by going out to distances where the Hubble flow is dominant. To make that leap secondary distance indicators of higher luminosity, (but often of lower precision and accuracy), were invoked. The secondary distance indicators specifically targeted by the Key Project for zero-point calibration by the Cepheids were the Tully-Fisher relation, the Surface Brightness Fluctuation method and the Fundamental Plane of galaxies, as well as two types of extremely bright explosive events, Type~I and Type~II supernovae. 

None of the secondary distance indicators have first-principles physics backing them up; they are largely empirical distances indicators.  Type Ia supernovae have most recently become the secondary indicator of choice because of 1) their brightness, which allows them to probe cosmological distances, 2) their standardizable maximum-light luminosities and 3) their low scatter in the Hubble diagram. At lower redshifts these candles are found to have absolute magnitudes with a dispersion of less than 5-6 percent per event \cite{burns_2020}. Establishing the absolute zero point of Type~Ia supernovae quickly became  the {\it de facto} standard means of deriving the local value of the expansion rate of the universe, with Cepheids providing the zero point calibration.

The final result from the HST Key Project, \ho = 72 $\pm$3 (stat) $\pm$ 7 (sys) \hounits,  was based on Cepheid distances to 31 galaxies, 18 of which were newly measured as part of the Key Project. The largest contribution to the systematic uncertainty (5\%), at that time, was that of the distance to the calibrating galaxy, the Large Magellanic Cloud (LMC), to which the distance had been measured using a wide variety of independent techniques. 

In what follows, we discuss in detail the two currently highest-precision methods for measuring distances to nearby galaxies, and for providing a tie-in to \sne: Cepheids and  the Tip of the Red Giant Branch (TRGB) method. For nearby galaxies, these two methods currently have the lowest measured scatter, their distances can be compared {\it galaxy by galaxy within the same galaxies}, and they  can be applied individually to samples of dozens of galaxies, in sharp contrast to other techniques at the moment.

We pay particular attention to systematic uncertainties, the essential issue in the measurement of galaxy distances, the determination of \ho, and for settling the question of whether there is additional physics beyond $\Lambda$CDM.

\section{Progress Since the Key Project: The Cepheid Distance Scale: 2001-2023}
\label{sec:cepheids}

Cepheids have held the place of being the gold standard for the measurement of extragalactic distances ever since Edwin Hubble's discovery of the expansion. A recent review of Cepheids as distance indicators is given by Freedman \& Madore \cite{freedman_madore_2023}. For more details on the nature of Cepheid variables themselves, the reader is also referred to some earlier reviews\cite{madore_freedman_1991, bono_2010, freedman_madore_2010, turner_2012}.

Following on the Key Project, Macri et al. \cite{macri_2001} obtained H band (F160W) observations of a subset of the Key Project galaxies using NICMOS on \hst. Their findings supported the assumption of universality for the extinction law for Cepheids: the VI photometry used in the Key Project Cepheid distance scale agreed with the augmented VIH distances employing the additional near-infrared observations.  This result suggested that there is no (extinction law) advantage in going to the extra effort to move the Cepheid calibration and its application into the IR. The study additionally showed that the lower spatial resolution in the H band imaging data led to more serious crowding effects than in the optical, an issue of even more concern as the sample of galaxies is augmented to include galaxies farther away.

\subsection{Chicago Carnegie Hubble Program (CCHP)}

The goal of the Chicago Carnegie Hubble Program (\cchp) is to increase the accuracy of measurements of \ho.  Initially begun 15 years ago (as the Carnegie Hubble Program), the program was  designed as a follow-up to the \hst Key Project, taking advantage of the mid-infrared capabilities of the Infrared Array Camera ($IRAC$) on $Spitzer$, and was undertaken in anticipation of the launches of \gaia and \jwst \cite{freedman_2012}. It followed up on \hst $NICMOS$ observations made in the F160W bandpass \cite{macri_2001} for Cepheids in 12 nearby galaxies, and the detailed JHK (complete lightcurve coverage) near-infrared, ground-based study of 92 Cepheids in the LMC\cite{persson_2004}.  Over time, the program was expanded to include not only Cepheids, but also TRGB \cite{freedman_2019, freedman_2021} and J-region Asymptotic Giant Branch (\jagb) stars \cite{madore_freedman_2020, freedman_madore_2020, lee_2021}, each of these being independent means of calibrating Type Ia supernovae (\sne) and thereby, \ho. The current focus of the \cchp is directed at exploiting the superb infrared sensitivity and high spatial resolution of the \jwst to improve the accuracy and precision of all three of these methods. 

\begin{figure}
    \centering
\includegraphics[width=\columnwidth]{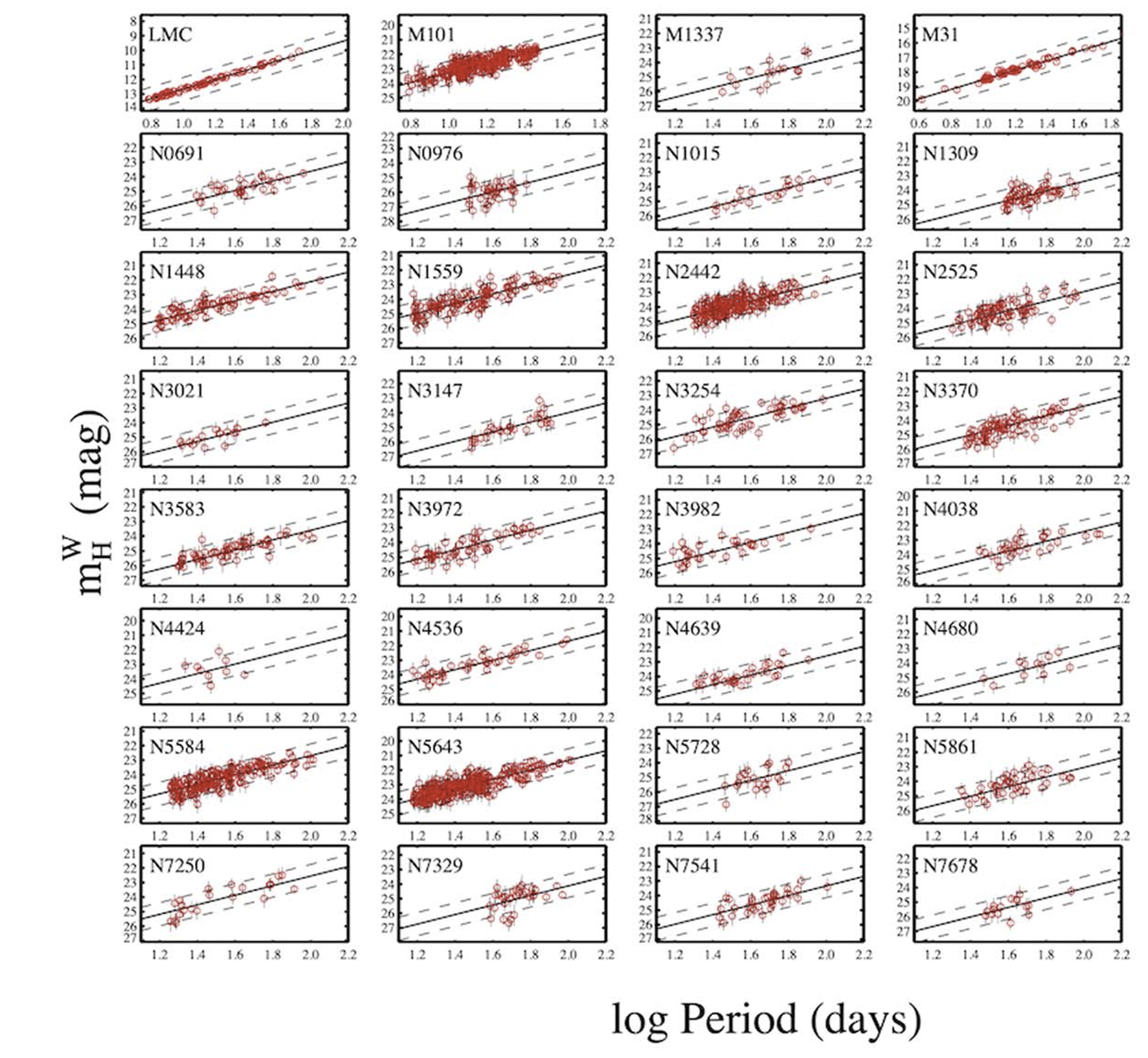}
    \caption{A sampling of reddening-free near-IR Wesenheit magnitude PL relations adapted from \cite{riess_2022}. Note the four-times larger scatter seen in virtually all of the SHoES galaxies compared to the fiducial scatter seen in the LMC and M31 (top row).}
\label{fig:SHOES1}
\end{figure}

\subsection{Supernova Ho for the Equation of State (SHoES)}
\label{sec:shoes}

The SHoES program \cite{riess_2022,Anderson_2023} has the goal of using \hstacs and \hstwfciii to extend and improve the Cepheid calibration of \sne for a measurement of \ho. Most recently, they have obtained  NIR observations of Cepheids in 42 \sn host galaxies with the aim of reducing the systematic uncertainties due to reddening and  metallicity.  Reddening corrections are obtained using a small number of (2 to 3) single-phase observations in the F814W and F555W bands from \hstacs. The distances are based primarily on  $\sim$6 low-signal-to-noise observations taken in the F160W ($H$) band. The resulting scatter in the F160W period--luminosity relations is typically of order $\pm$0.4 -- 0.5~mag,
which is about a factor of four times greater than the intrinsic dispersion observed in the uncrowded sample of Cepheids in the LMC, for instance. The zero-point calibration is set by  Early Data Release 3 (EDR3) geometric parallaxes, masers in the galaxy NGC 4258, and detached eclipsing binaries in the LMC. Their most recent result quotes a 1\% uncertainty with \ho = 73.04 $\pm 1.04$ \hounits, based on their sample of 42 galaxies with
distances in the range from 7 out to 80 Mpc.

\section{Tip of the Red Giant Branch (TRGB) Distance Scale: 1993-2023}
\label{sec:trgb}

The TRGB provides one of the most precise and accurate means of measuring distances in the local universe. Observed color-magnitude diagrams of the Population~II stars in halos of  nearby galaxies reveal a sharp discontinuity in the red giant branch (RGB) luminosity function at a well-determined magnitude. This feature is easily identified and corresponds to the core helium-flash luminosity at the end phase of RGB evolution for low-mass stars. As a result, the TRGB provides a superb standard candle in the I band \citep{lee_1993, rizzi_2007, salaris_2002, madore_2009, freedman_2019, jang_2021}, and it is a standardizable candle in the near infrared \cite{dalcanton_2012, wu_2014, madore_2018, durbin_2020}.  The method is described in more detail in a number of reviews \cite{madore_freedman_1999,freedman_madore_2010,madore_freedman_2023}.

In brief, the underlying theory for why the TRGB is an excellent standard candle is well-developed \citep{salaris_1997, Salaris_2005, bildsten_2012,kippenhahn_2013, serenelli_2017}. For low-mass stars with masses $M\lesssim 2 M_{\odot}$, their evolution ascending  the red giant branch consists of a shell that is burning hydrogen immediately above  a degenerate helium core. The mass of the helium core increases with freshly formed helium  from the shell burning, until the core mass reaches a threshold value of about 0.5 M$_{\odot}$ {\bf independent of the initial mass of the star}. At this stage the core will have reached a temperature of about 10$^8$ degrees, at which point the triple-alpha process (helium burning) can commence. Because the core is degenerate and cannot expand, a thermonuclear runaway ensues,  injecting  energy  that overcomes the core degeneracy, and changing the equation of state. 
The star then rapidly evolves off the red giant branch  to the (lower-luminosity) horizontal branch or the red clump, thereafter undergoing sustained core helium burning.

 \begin{figure}
    \centering
\includegraphics[width=\columnwidth]{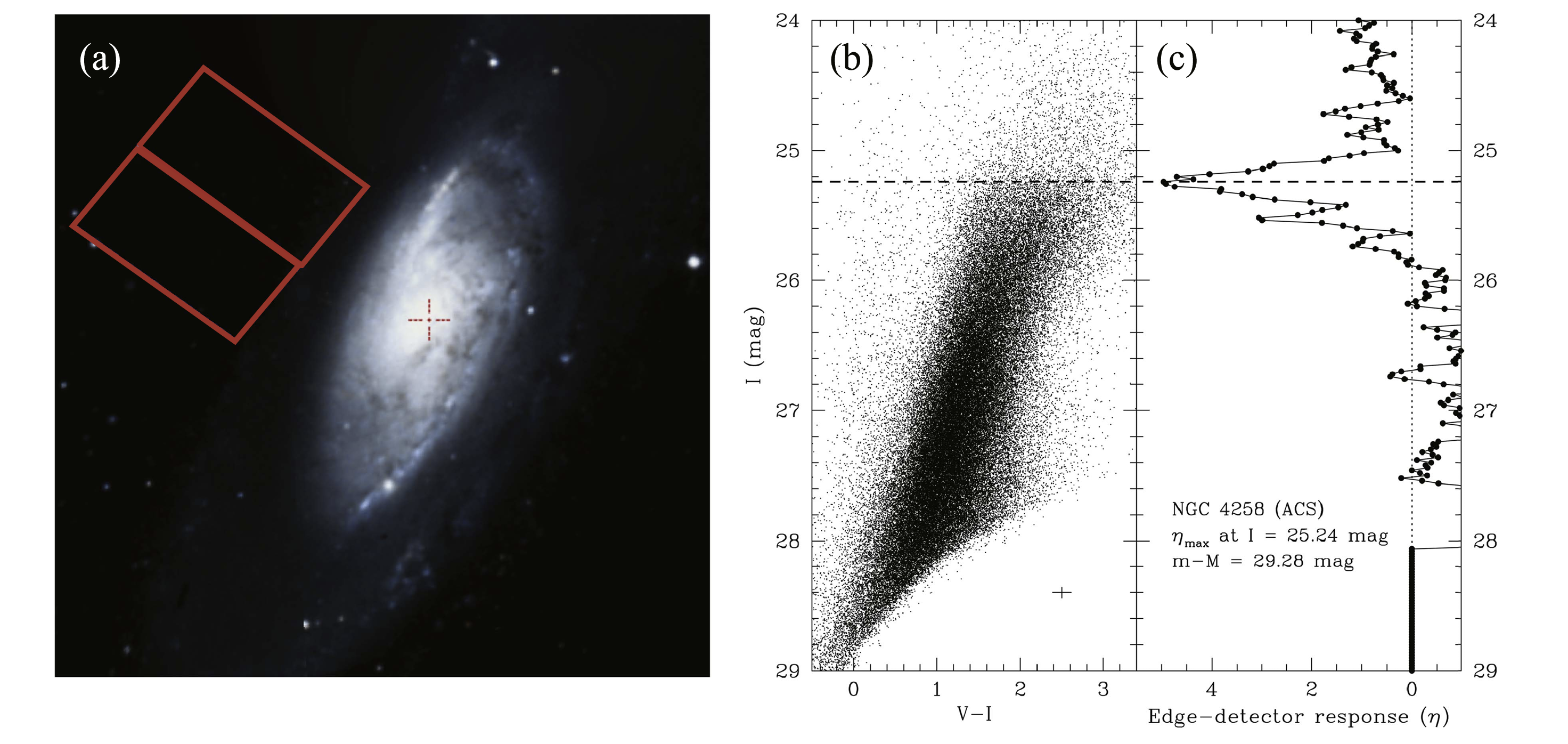}
    \caption{Left Panel -- An example of a  halo field chosen to be along the minor axis of the galaxy \ngc 4258. Right Panel -- The I-band vs (V-I) color-magnitude diagram for the RGB stars detected in the halo of \ngc 4258. To the far right is the Sobel filter edge-detector response function applied to the RGB luminosity function. The peak in the edge detector indicates the discontinuity defining the TRGB. Adapted from \cite{Mager_2008}.}
\label{fig:TRGB1}
\end{figure}

\vfil\eject
The TRGB method has been used widely for the determination of distances to galaxies of various types in the local universe. The application of the TRGB method  far exceeds the number of measurements of Cepheid distances\footnote{Approximately 1,000 TRGB distances to about 300 galaxies are compiled in NED; less than 70 galaxies have Cepheid distances to date.}. The reason is practical:  Cepheids are variable stars requiring  observations at many epochs to determine periods, amplitudes and light curves for the construction of time-averaged period-luminosity relations. In contrast, TRGB stars are non-variable and have constant I-band magnitudes as a function of color and metallicity, requiring only a single-epoch observation.  In addition, TRGB stars can be observed in galaxies of all morphological types, whereas Cepheids are present only in late-type galaxies.

\subsection{Chicago Carnegie Hubble Program (CCHP) and the TRGB}
\label{sec:CCHP+TRGB}

One of the primary goals of the Chicago Carnegie Hubble Program (\cchp) is  to pursue an alternative route to the calibration of \sne and thereby provide an independent determination of \ho via measurements of the TRGB in nearby galaxies. This method has a precision equal to or better than the Cepheid Leavitt law, and its current accuracy is also comparable. 
The calibration of the zero point of the I-band TRGB method and its application to the extragalactic distance scale has recently been reviewed by Freedman \cite{freedman_2021}. 

Freedman et al. \cite{freedman_2019}  presented a determination of \ho based on TRGB distances to 15 galaxies that were hosts to 18 Type Ia supernovae (\sne). The \hstacs fields were selected to target the halos of the galaxies where the effects of dust are minimal, and, at the same time,   to specifically avoid contamination by younger and brighter disk asymptotic giant branch (AGB) stars. This calibration was then applied to a sample of 99 significantly more distant \sne that were observed as part of the Carnegie Supernova Project (CSP)\cite{krisciunas_2017}.  The calibration has been updated \cite{freedman_2020, freedman_2021}, and is currently based on our independent calibrations of the TRGB absolute magnitude that are internally self consistent at the 1\% level. The method yields a value of \ho = 69.8 $\pm $ 0.6 (stat) $\pm$ 1.6 (sys) \hounits.  This value differs only at the 1.2$\sigma$  level from the most recent Planck Collaboration \cite{planck_2018} value of \ho. It is smaller than previous estimates of the Cepheid calibration of SNe Ia \cite{freedman_2012, riess_2022} but still agrees well, at better than the 2$\sigma$ level. Alternatively, adopting the \sne catalog from the \shoes collaboration \cite{scolnic_2015} results in little change with \ho = 70.4 $\pm$ 1.4 $\pm$ 1.6 \hounits \cite{freedman_2019}.

\subsection{Other Determinations of the Hubble Constant based on the TRGB}

Recently members of the \shoes collaboration \cite{scolnic_2023} have undertaken to provide an `optimized unsupervised algorithm' (called CATS) to measure TRGB distances and determine \ho. They find a value of \ho =  73.22  $\pm$ 2.06 \hounits,  apparently in better agreement with the Cepheid calibration. However, it can be easily demonstrated that this approach currently contains serious flaws.

1) Take, for example,  two of the galaxies in their sample, \ngc~4038 and \ngc~4536.  The TRGB distances that their unsupervised algorithm gives the highest weight to (they allow for several TRGB distances to an individual galaxy and employ various means of combining them) are significantly closer than their own published \shoes Cepheid distance measurements \cite{riess_2022} by 0.7 and 0.6 mag, that is 30\% and 40\% offsets in distance, respectively, ultimately pulling \ho to higher values. These TRGB distances are the ones that their unsupervised method ranks as ``better measurements''. This result stands in stark contrast to the excellent agreement between the published  Cepheid distances  in Riess et al. \cite{riess_2022} and TRGB distances in Freedman et al. \cite{freedman_2019}, which in the mean, agree to 0.007 mag.

2) In many cases, the unsupervised method mistakes the well-known asymptotic giant branch (AGB) for the RGB (e.g., \ngc 4038). This is a documented problem that has been addressed by many authors previously in the literature \cite{schweizer_2008, jang_lee_2015, MF_23}.

\begin{figure}
    \centering
\includegraphics[width=9cm]{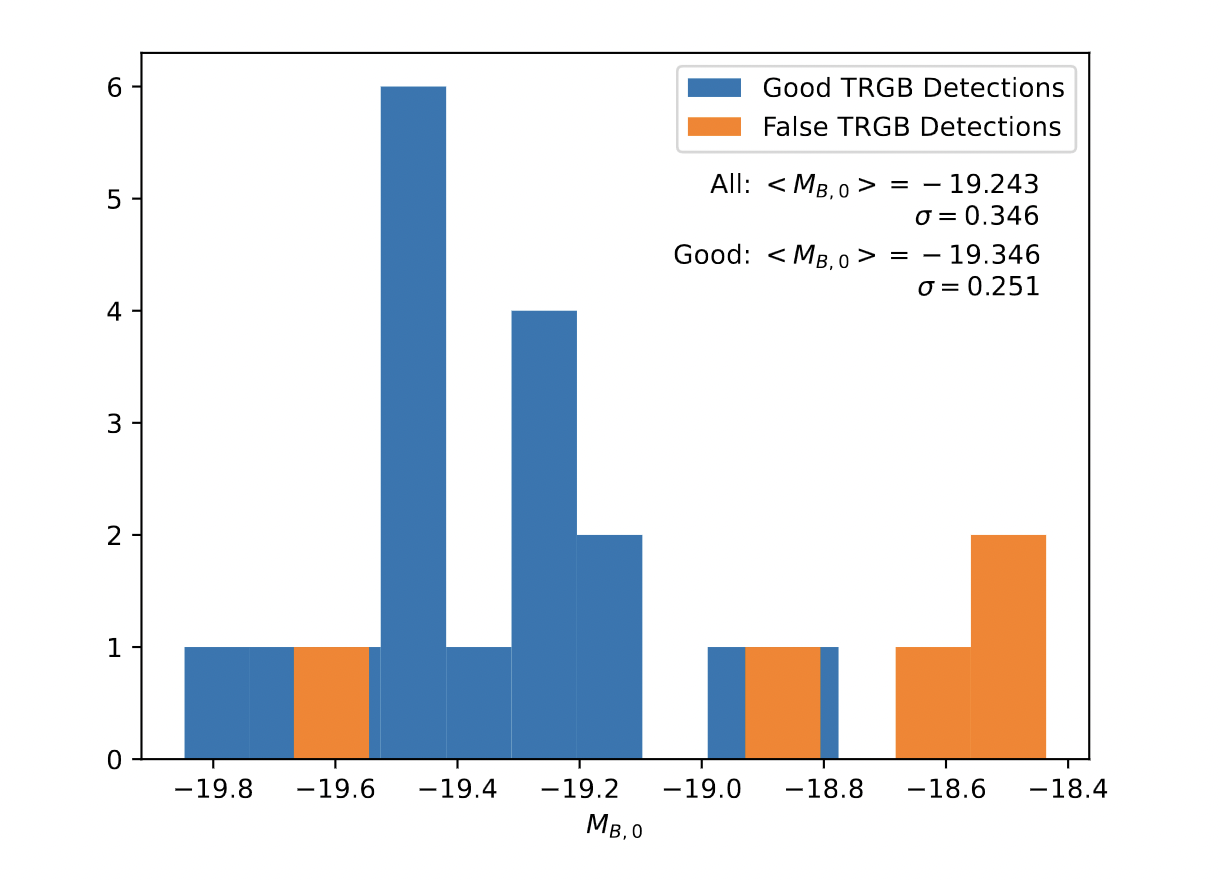}
    \caption{Histogram of the peak \sne magnitudes from Tables 1 and 2 and Equation 3 of Scolnic et al. \cite{scolnic_2023} (Hoyt, private comm.) 
    The cases where the AGB has been mistakenly identified for the RGB are shown in orange. The dispersion in the \sne peak magnitudes is erroneously increased as a result of these anomalously fainter magnitudes, biasing the result and leading to a higher value of \ho.}
\label{fig:Taylor_sn}
\end{figure}

3) That these distances are problematic is additionally demonstrated by the fact that in adopting  their unsupervised TRGB distances, the $rms$ dispersion in the \sne peak magnitude for the TRGB \sne host galaxies increases from $\pm$0.12 mag \cite{freedman_2019} to $\pm$0.346 mag (see Figure \ref{fig:Taylor_sn}, Hoyt private communication). Their re-analysis increases the scatter in the \sne peak magnitudes to a level that is a more than a factor of three greater than that seen in the distant \sne, $\pm$0.10 \cite{burns_2018}.

Although an algorithmic method might be desirable in future studies, it is unambiguously clear that this current unsupervised method itself necessitates supervision, and the results cannot be used to claim that this approach brings the Cepheid and TRGB distance scales into better agreement. 

\begin{figure}
    \centering
\includegraphics[width=9cm]{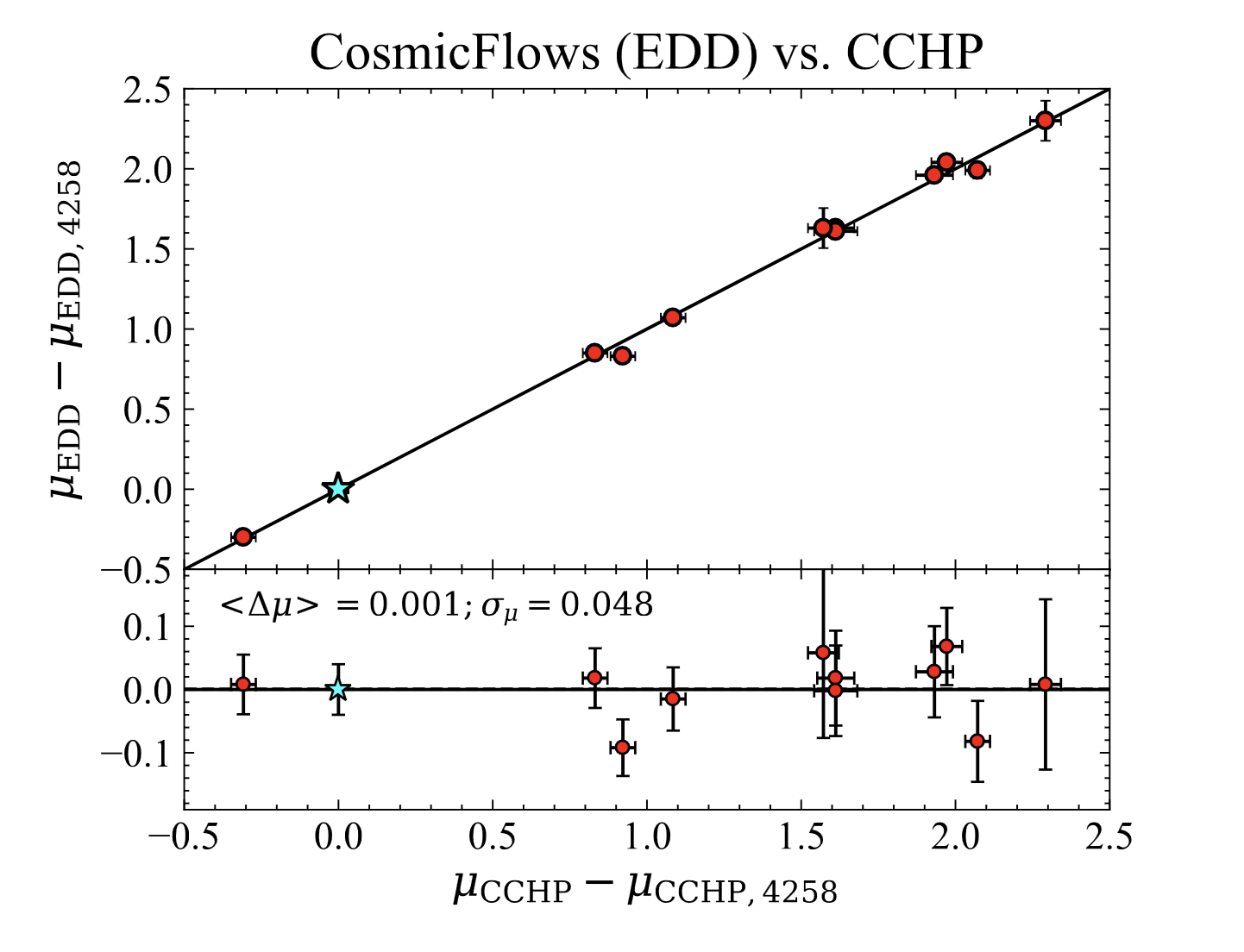}
    \caption{A comparison of TRGB distances from the EDD \cite{anand_2021} and the CCHP \cite{jang_lee_2017a, freedman_2019, jang_2021, hoyt_2021b}. The distances are calibrated relative to \ngc 4258 (blue star). A line of unit slope is shown in the top panel. In the bottom panel, the median offset value is shown (dashed line), as well as at zero offset (solid line). These two independent analyses show excellent agreement.} 
\label{fig:CCHP+EDD}
\end{figure}

Comparisons of the CCHP TRGB distances with those in the Extragalactic Distance Database (EDD) were undertaken by \cite{hoyt_2021b} and \cite{anand_2021}. These comparisons provide an important  external check of the TRGB method since different approaches to the the analysis were taken, and carried out completely independently by separate research groups. For example EDD used DOLPHOT and applied a maximum likelihood fitting method whereas the CCHP used DAOPHOT and an edge-detection (Sobel filter) algorithm. Adopting a consistent  \ngc 4258 calibration, the difference is only  
0.001 $\pm$0.048 mag \cite{hoyt_2021b} (see Figure \ref{fig:CCHP+EDD}); i.e., the analyses and the relative distances agree remarkably well. A remaining discrepancy is the absolute calibration of the TRGB at the 0.06~mag level (see \cite{freedman_2021} and \cite{anand_2021}, line 1, Table 4). This difference results from the choice to calibrate either in the outer halo (CCHP) or the disk of \ngc 4258 (EDD). The outer halo provides a dust-free and uncrowded environment. 

\section{Anchors to the Distance Scale}

At present the overall accuracy in the determination of \ho is limited by the small number of galaxies that `anchor' the Cepheid and TRGB distance scales; that is, galaxies for which there are geometric distances, acting as the first stepping stones out to the more distant galaxies. In the case of the Cepheid distance scale, there are only three such anchors: the Milky Way, the Large Magellanic Cloud (LMC) and the maser galaxy \ngc~4258. In the case of the TRGB, there is one additional anchor, the Small Magellanic Cloud (SMC). JAGB stars also have the Milky Way, LMC, SMC and NGC~4258 as anchors.

\subsection{Large Magellanic Cloud (LMC)}

At the conclusion of the Key Project, the largest component of the systematic error budget was the contribution from the adopted uncertainty to the distance of the LMC. A distance modulus to the LMC of 18.5~mag  was adopted, with a very conservative uncertainty of $\pm$ 0.1 magnitudes, reflecting the wide range of published distance moduli at the time (18.1 to 18.7 mag) \cite{freedman_2001}. 

The distance modulus to the LMC has been improved significantly since the time of the Key Project,  based on measurements of 20 detached
eclipsing binary (DEB) stars in the LMC \cite{pietrzynski_2019}. This method gives a distance modulus of 18.477 $\pm$ 0.004 (stat) $\pm$ 0.026 (sys), corresponding to a distance uncertainty of only 1.2\%. The DEB value is in exact agreement with measurements of the Cepheid Leavitt law based on 3.6 $\mu$m mid-infrared measurements from the Spitzer Space Telescope \cite{monson_2012, scowcroft_2011}. Furthermore this value is only 0.023~mag different from the Key Project value, meaning that the LMC zero-point calibration adopted at that juncture has withstood the test of time, at a $\sim$1\% level of accuracy.

\subsection{Milky Way Parallaxes: Hipparcos, HST and Gaia}

There have also been enormous gains in the measurement of parallaxes to Cepheids in the Milky Way in the past 20 years, from {\it Hipparcos} \cite{perryman_2001} to \hst and its Fine Guidance Sensor \cite{benedict_2002, benedict_2007, benedict_2011} (which provided the calibration for the Spitzer Cepheid PL relation \cite{freedman_2012}), culminating most recently with measurements from \gaia\cite{gaia_lindegren_2021a,gaia_lindegren_2021b}. The \gaia measurements are revolutionizing studies of the Milky Way; for example, see \cite{cantat-gaudin_2022}.

The \textit{Gaia} Early Data Release 3 (EDR3)  database \cite{gaia_brown_2021} contains parallaxes, proper motions, positions and photometry  for 1.8 billion sources brighter than G = 21~mag \cite{gaia_lindegren_2021b}. At the end of its mission, \gaia is expected to provide astrometry reaching  tens of microarcsecond accuracy. For Milky Way Cepheids, TRGB stars and other distance indicators, this level of accuracy will ultimately set the absolute calibration to an accuracy of  $<$1\%, an accuracy critical for helping to resolve the \ho tension. However, this challenging high accuracy has not yet been achieved owing to a zero-point offset \cite{lindegren_2016}  resulting from the fact that the basic angle between the two \gaia telescopes is varying. There is  a variance in the parallaxes (the systematic uncertainty  measured relative to the background-quasar reference frame, defined by 550,000 quasars in the International Celestial Reference System) and a zero-point offset  of  --17 $\mu$as (in the sense that the \gaia parallaxes are too small).  Unfortunately this offset results in a degeneracy with the absolute parallax, and is limiting the ultimate accuracy required to reach the 1\% target. In addition, these variations lead to zero-point corrections that are a function of the magnitude, color, and position of the star on the sky \cite{lindegren_2018, arenou_2018}. The \gaia Collaboration has emphasized \cite{bailer-jones_2021, fabricius_2021} that not only is there a significant variance in these measured  offsets over the sky, but the EDR3 uncertainties in the parallaxes  for different objects are correlated as a function of their angular separations \cite{ gaia_lindegren_2021a, gaia_lindegren_2021b}. 

Furthermore \gaia EDR3 parallaxes uncertainties have also been shown to be underestimated \cite{fabricius_2021}, with the `unit weight' uncertainties of the catalog (the factors by which the formal errors need to be increased to reflect the actual level of uncertainty) having a multiplicative factor of $\sim$1.2 for the majority of stars, but in some instances rising to a factor of more than 2. Unfortunately, the most significant underestimates occur for brighter stars \citep{el-badry_2021}, including  the magnitude range over which many of the Milky Way field Cepheids lie. Of additional and serious  concern for the Cepheid distance scale, the parallax offset adopted turns out to be degenerate with the metallicity coefficient adopted \cite{owens_2022}, which together with the uncertainties in the measured parallaxes, leads to a systematic floor at the 4\% level. With the exception of \cite{riess_2022} these studies agree that a 1\% calibration based on \gaia parallaxes has not yet been established.

\subsection{NGC 4258}
\label{sec:maser_anchor}

The nearby  spiral galaxy   \ngc 4258, at a distance of 7.6 Mpc, provides an additional anchor or zero-point calibration for the local distance scale. This galaxy is host to a sample of H$_2$O megamasers within an accretion disk that is rotating about a supermassive black hole, from which a geometric distance to the galaxy has been measured \citep{humphreys_2013, reid_2019}. (For more details on the method, see Section \ref{sec:masers}.) The  geometric distance modulus measured most recently to \ngc 4258 is $\mu_o$ = 29.397 $\pm$ 0.033 mag  \cite{reid_2019}, a 1.5\% measurement. 

As a consistency check, the distance to \ngc~4258  can be determined based on \hst measurements of the TRGB in its outer halo,  calibrated by the LMC  \cite{hoyt_2023a}.  Adopting the measured apparent TRGB magnitude of m$_{o~F814W}^{~N4258}= 25.347 \pm 0.014 \pm 0.005$ \citep{jang_2021}, results in a distance modulus of $\mu_o$ =  29.392 $\pm$ 0.018 $\pm$ 0.032 mag that agrees with the maser distance modulus of 29.397 $\pm$ 0.033 mag at a level of better than 1\%  ($<$0.2$\sigma$).  

The Cepheid calibration, however, does not yield as good agreement with that of the maser distance, and ultimately depends on the sensitivity of Cepheid luminosities to metallicity.  A calibration of the Cepheid distance to \ngc 4258 based on the LMC differs from the maser distance by  2.0-3.5$\sigma$, adopting different published slopes for the metallicity correction \cite{efstathiou_2020}. However, since the Milky Way and \ngc~4258 metallicities are very similar, a calibration of \ngc 4258 based on the Milky Way should be independent of a metallicity effect. Yet, if the Milky Way is adopted as the anchor galaxy to determine the Cepheid distance to \ngc~4258, a distance modulus of 29.242 $\pm$ 0.052 is obtained, which differs from the maser distance by 7\% at a 2$\sigma$ level of significance. These kinds of differences in the anchors to the distance scale are very important to resolve in the context of assuring that a 1\% \ho value is in hand. 

\section{Type Ia Supernovae}

The numbers of well-observed \sne  useful for measuring \ho has continued to grow with time \cite{brout_2022}. These include the nearby \sne out to distances of $\sim$30-40 Mpc that can be calibrated using \hst distances from the TRGB or Cepheids. If systematic effects due to crowding can be established to be small (but see Section \ref{sec:jwst_cepheids} below), perhaps the calibration can be reliably extended to $\gtrsim$50 Mpc. The \shoes collaboration now has 42 galaxies for which Cepheids have been discovered, out to a distance of 80 Mpc. The nearby \sne that can be observed with \hst that occur in galaxies for which the TRGB or Cepheids can be measured typically occur only about once per year \cite{riess_2022}. 

We discuss below two programs that currently calibrate the Cepheid and TRGB distance scales: the Carnegie Supernova Project (CSP) and Pantheon+.

\subsection{Carnegie Supernova Project (CSP)}

The goal of the CSP was to provide a homogeneous, intensive, high-cadence, multi-wavelength ($uBVgriYJH$) follow-up of nearby \sne and SN II \cite{contreras_2010}. Not a survey program, the idea was to obtain a consistent data set with careful attention to photometric precision and systematics, critical for applications to cosmology, as well as for studying the physical properties of the supernovae themselves\footnote{The CSP data are available at http://csp.obs.carnegiescience.edu/data.}. The program utilized a fixed set of instruments, photometric standard stars, and instrumental reduction procedures, catching most of the supernovae well before maximum, and with high signal to noise, avoiding the challenges otherwise faced in minimizing systematic differences between multiple data sets/instruments/etc. \cite{krisciunas_2017}. Optical spectra were also obtained with high cadence \cite{folatelli_2013}. The bulk of the observations were carried out at Las Campanas Observatory using the 1-m Swope and 2.5-m du Pont telescopes. The first part of the CSP (CSP-I) was carried out from 2004-2009. A second phase of the CSP (CSP-II) was carried out from 2011-2015, and was optimized for the near-infrared \cite{phillips_2019, hsiao_2019}.

The reduction of the CSP light-curve photometry was undertaken using an analysis package called  SNooPy \cite{burns_2018}. The Hubble diagram for the CSP-I \sne sample, calibrated by Cepheid distances from \cite{riess_2016} was presented in  \cite{burns_2018}. These authors found a value of \ho = 73.2 $\pm$ 2.3 \hounits based on H-band data; and a value of \ho = 72.7 $\pm$ 2.1 \hounits using B-band data. A TRGB calibration of the CSP-I sample was given by \cite{freedman_2019} and updated in \cite{freedman_2020, freedman_2021}. As discussed in Section \ref{sec:CCHP+TRGB} above, the TRGB calibration gives a slightly lower value of \ho = 69.8 $\pm $ 0.6 (stat) $\pm$ 1.6 (sys) \hounits. 

Recently \cite{uddin_2023} have used the \sne data from the CSP-I and II (an increase by a factor of three in the numbers of \sne over CSP-I alone) to calibrate the Cepheid distance scale, as well as the TRGB (and Surface Brightness Fluctuations, a secondary distance indicator). Using B-band light-curve fits, they find \ho = 73.38 $\pm$ 0.73 \hounits based on a calibration of Cepheids. For the TRGB calibration, they find \ho = 69.88 $\pm$ 0.76 \hounits, both in good agreement with previously published Cepheid and TRGB studies. They conclude that the differences amongst the various calibrators can be explained as a result of systematic errors, and that taking these into account removes the existing \ho tension (see also Section \ref{sec:crisis}). 

\subsection{Pantheon+}

The Pantheon+ analysis \cite{scolnic_2022} currently consists of 1550 individual \sne, superseding earlier Pantheon \cite{scolnic_2018} and Joint Light-Curve \cite{betoule_2014} analyses.  The analysis knits together and standardizes the B-band photometry from 18 individual surveys obtained with a wide variety of telescopes and instruments\footnote{ The Pantheon+ catalog is available at https://github.com/PantheonPlusSH0ES/DataRelease.}. The sample includes \sne in the redshift range 0 $<$ z $<$ 2.3; the subset used for constraining \ho are those for which 0.023 $<$ z $<$ 0.15.

The \shoes Cepheid calibration of the Pantheon+ \sne sample from \cite{scolnic_2022} results in a value of \ho = 73.04 $\pm$ 1.04 \hounits for the 277 \sne with 0.023 $<$ z $<$ 0.15, as noted previously in Section \ref{sec:shoes}.

Ultimately, it is expected that the \sne samples will continue to grow as future large-scale (and homogeneous) surveys like the Legacy Survey of Space and Time (LSST) \cite{Ivezic_2019} and the Nancy Grace Roman Space Telescope \cite{hounsell_2018} become available.

\section{J-Region Asymptotic Giant Branch (JAGB) Distance Scale:  2000-2023}
\label{sec:jagb}

\begin{figure}
\centering
\includegraphics[width=8.0cm]{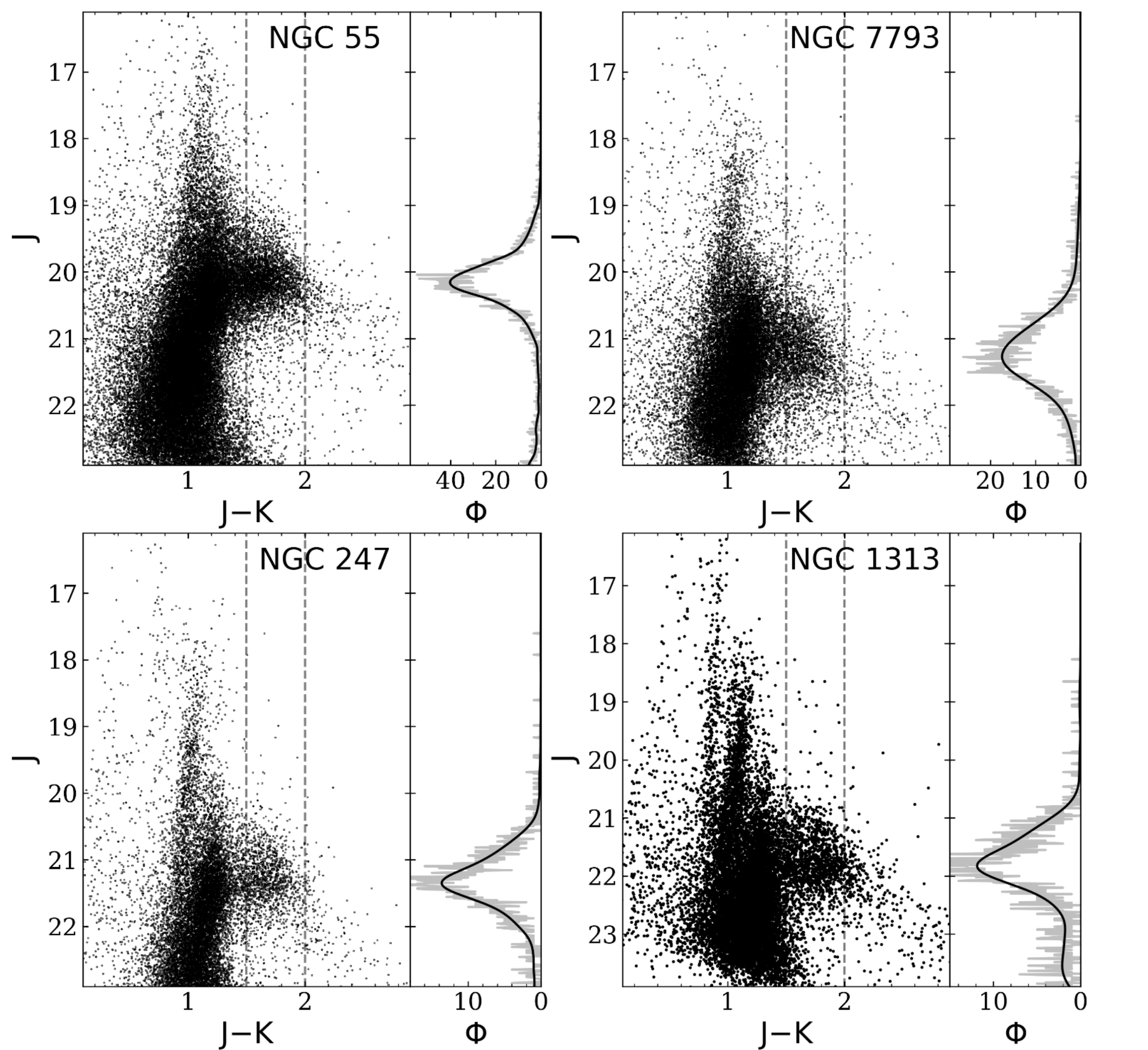}
\caption{Ground-based (Magellan/FourStar) near-infrared CMDs of four nearby galaxies observed by Lee et al. (in preparation), illustrating the well defined, single peaked J-band luminosity functions characteristic of the JAGB population in the color range 1.5 $<$ (J-K) $<$ 2.0.}
\label{fig:jagb_lee2}
\end{figure}

The JAGB method is emerging as one of the most promising methods for measuring the distances to galaxies in the local universe. JAGB stars were first identified as a distinct class of objects in the LMC \cite{nikolaev_weinberg_2000, weinberg_nikolaev_2001}, demonstrated to be very high precision distance indicators, and then successfully used to map out the  back-to-front geometry of the LMC. Two decades later the method was applied in an extragalactic distance scale context \cite{madore_freedman_2020, freedman_madore_2020, ripoche_2020}. Together, these  studies have demonstrated that there is a well-defined class of carbon stars with a nearly constant luminosity in the near-infrared; i.e., an excellent standard candle for distance measurements. These (thermally-pulsating AGB) stars have a low intrinsic dispersion, specifically in the near-infrared J band, of only $\pm$0.2 mag \cite{nikolaev_weinberg_2000}, and they can be identified on the basis of their near-infrared colors alone, being distinguished from bluer O-rich AGB stars, as well as being segregated from redder, extreme carbon stars (see Figure \ref{fig:jagb_lee2}). 

\begin{figure}
\centering
\includegraphics[width=7.0cm]{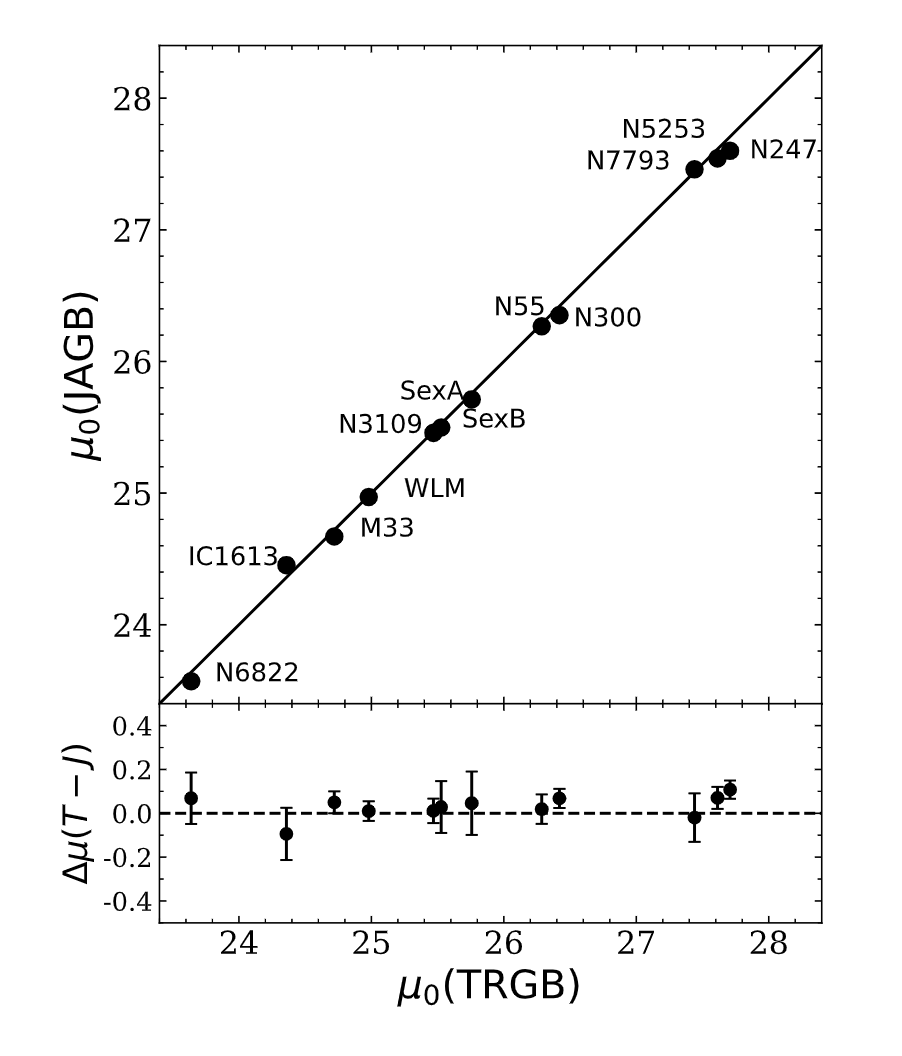}
\caption{Comparison of ground-based JAGB and TRGB distance moduli to a dozen nearby galaxies (Lee et al. in preparation). The lower panel shows the magnified differences between the moduli which have a combined scatter of only $\pm$0.06~mag. For this sample of galaxies this scatter puts upper limits (of a few percent) on the impact of metallicity differences, differential internal reddening and potential star formation history differences between these galaxies.}
\label{fig:jagb_lee1}
\end{figure}

Freedman \& Madore (2020) measured JAGB carbon-star distances to a sample of 14 galaxies out to ~27 Mpc, calibrated using the LMC and the SMC, and compared them to previously published distances using the TRGB. They found that the distance moduli agreed extremely well (at the 1\% level), with a (combined) scatter amounting to only $\pm$4\%.  The good agreement with the TRGB distances suggests that the effects of metallicity for this well-defined color-range of carbon stars are small. A number of additional extensive tests of this method have recently been carried out by Lee and collaborators \cite{lee_2021, lee_2022, lee_2023a} as well as Zgirski et al.~\cite{zgirski_2021} in several nearby galaxies, confirming the excellent agreement with distances measured with the TRGB and Cepheid distance scales, and again indicating that metallicity and star formation effects are small (see Figure \ref{fig:jagb_lee1}).

Recent modeling of AGB star evolution has been carried out by many authors \cite{marigo_2017,  salaris_2014, pastorelli_2020}. Significant challenges remain in the detailed modeling (e.g., treatment of convection, overshoot, winds and mass loss), but the broad outlines are well-characterized.  A carbon star is defined such that the atmosphere contains more carbon than oxygen; i.e., a ratio of C/O $>$1. The path to becoming a carbon star occurs during the thermally pulsing evolutionary phase for AGB stars. As a result, carbon can be brought to the surface, particularly  during the third and later (generally deeper) dredge-up phases \cite{iben_renzini_1983, herwig_2013, habing_olofsson_2004}.  For stars with solar metallicity, recent studies conclude that the initial mass for carbon-star formation is between 1.5 and 3.0 to 4.0 M\textsubscript{\(\odot\)} \cite{Marigo_2022}, with a similar range for stars with Z = 0.008 \cite{pastorelli_2020}. 

The reason for the well-constrained luminosity of carbon stars is two-fold: (1) younger, more massive (hotter) AGB stars burn their carbon at the bottom of the convective envelope before it can reach the surface of the star \cite{boothroyd_1993}, whereas (2) for the oldest, less massive AGB stars, there is no third, deep dredge-up phase. Thus, carbon stars are formed only in the intermediate mass range where carbon-rich material can both be dredged up and survives so that it can be mixed into the outer envelope. 

In summary, the JAGB method offers a number of advantages for distance measurement, as previously enumerated \cite{freedman_madore_2020}. (1) 
They  are  easily identified by their colors and magnitudes in the infrared. (2) They have a low intrinsic dispersion in the J band of only $\pm$0.2 mag. (3) They are about one magnitude brighter than those defining the TRGB. (4) They are found in all galaxies that have intermediate-age populations, and the JAGB method is, therefore, applicable to a wide range of galaxy types. (5) Near-infrared observations offer the advantage of reduced intrinsic variability and reduced reddening. (6) No multi-epoch observations are required to determine periods as, for example, is the case for Cepheid and Mira variables; observations of JAGB stars in two infrared bands, at a single epoch, are all that is needed.

With further development, testing and application the JAGB method has the potential to provide an independent calibration of Type Ia supernovae (\sne), especially with $JWST$. JAGB stars are brighter than the TRGB and thus can be detected at greater distances, allowing greater numbers of calibrating galaxies for the determination of \ho. As is the case for the TRGB and Cepheids, JAGB stars are amenable to theoretical understanding and further improved empirical calibration. Early tests show little dependence, if any, of the JAGB magnitude with metallicity of the parent galaxy (see Lee et al. \cite{lee_2023a} and Figure 9), and therefore suggest that the JAGB method has considerable promise for providing high-precision distances to galaxies in the local universe that are largely independent of distances derived from the Leavitt Law and/or the TRGB method.

\section{Other Methods}
\label{sec:other}

\subsection{Surface Brightness Fluctuations (SBF)}

For most distance indicators crowding of individual stars by the surrounding population of stars is a major source of systematic uncertainty; a systematic that increases in its effects as the targets being measured are found at increasing distances. Thirty-five years ago Tonry \& Schneider \cite{tonry_schneider_1988} introduced a novel technique, called the Surface-Brightness Fluctuation (SBF) method that takes crowding (a systematic effect that depends on distance) and  turns a quantitative measure of the crowding into a means of measuring distances. The method has recently been extensively reviewed in  \cite{cantiello_blakeslee_2023}.

The SBF method applies best to elliptical galaxies, and with caution, to the bulges of bright, early-type spiral galaxies, where the effects of dust and recent star formation can be mostly avoided. At a given surface brightness (which is by definition independent of distance) the degree of crowding of any pre-specified population of stars will increase/degrade with distance as the mean separation  of those same stars also decreases inversely with distance. A measure of the observed granularity in the image, which is  used to determine a distance, is found in the power spectrum of the targeted field of view. 

Recent applications of the SBF method \cite{blakeslee_2021, khetan_2021} have led to values of  \ho = 73.3 $\pm$ 0.7 (stat) $\pm$ 2.4 (sys) and \ho = 70.50 $\pm$ 2.37 (stat) $\pm$ 3.38 (sys)\hounits. The most important error terms \cite{cantiello_blakeslee_2023}  are (i) sky background subtraction [0.02 mag], (ii) characterization of the point spread function [0.03 mag], (iii) details of the power spectrum fitting [0.02 mag], (iv) residual variance in the power spectrum, due to globular clusters and background galaxies too faint to be detected and masked directly [0.05 mag], and (v) extinction. Values in square brackets  are the errors due to these terms as estimated by \cite{cantiello_blakeslee_2023} Section 1.4.1.

Finally, it should be noted that the SBF method is a secondary distance indicator (as are other notable examples, including Type~Ia supernovae and the Tully-Fisher relation) given that it is not calibrated from first principles, nor is it calibrated from geometric/parallax methods. Rather, SBF  is currently being calibrated using (primarily) Cepheid and (a small number of) TRGB distances to galaxies close enough for those methods to provide a tie-in. 

The very strong intrinsic-color dependence of the SBF characteristic magnitude is assumed to be due to the effects of the metallicity distribution on the RGB colors, in combination with differing contributions of AGB stars due to different star formation histories. Uncrowded, high signal-to-noise color-magnitude diagrams of the stellar populations underwriting the SBF method would be important to have for a range of integrated colors in nearby elliptical galaxies so as to quantitatively constrain any potential systematic effects.

With \jwst/\nircam and other upcoming facilities, it will be possible to surmount the current 100 Mpc distance limit for SBF distances, perhaps taking it out to 300~Mpc, thus reducing the uncertainty from peculiar motions, as well as improving the statistical precision. 

\subsection{Masers}
\label{sec:masers}

H$_2$O mega-masers provide a powerful geometric tool for measuring extragalactic distances. These astrophysical masers, often found in the accretion disks around supermassive black holes, are akin to lasers, instead operating in the microwave regime. Water molecules in these disks amplify background radiation and produce coherent emission. The radial velocity shifts exhibited by the megamaser sources, observed with high-resolution radio interferometry, allow for the detailed mapping of the rotational dynamics of the maser-bearing accretion disk. By applying Kepler's laws to the derived rotation curve, the mass of the central supermassive black hole can be determined. A direct geometric distance to the galaxy can be obtained making use of the  constrained orbital dynamics and precise angular measurements provided by Very Long Baseline Interferometry (VLBI) \cite{lo_2005}.  Allowing for warps and radial structure, the approximately Keplerian rotation curve for the disk can be  modeled. The nearest and best-studied galaxy, NGC~4258, at a distance of about 7.5~Mpc, is too close to provide an independent measurement of the Hubble constant (i.e., free from local velocity-field perturbations) but it serves as a geometric anchor for the distance scale.

The {\it Megamaser Cosmology Project} has measured maser distances to 6 galaxies within 130~Mpc \cite{pesce_2020}. Adopting an average peculiar velocity uncertainty of $\pm$250~km/s they determine a value of  \ho = 73.9 $\pm$ 3.0 \hounits, with a range of  values spanning 71.8 to 76.9 \hounits, allowing for different means of correcting for peculiar velocities. Sadly, the numbers of galaxies for which this technique can be applied turns out to be very small; hence, it will never rival, for example, \sne (for which there are upwards of 1,000 host galaxies)  in statistical precision. 

\subsection{Strong Gravitational Lensing}

Strong gravitational lensing offers an independent route for determining \ho  with the advantage that it can be carried out at cosmological distances (a one-step method), providing crucial cross-checks against measurements of the local distance scale and CMB measurements. In a gravitational lensing event, a massive foreground object (like a galaxy cluster) distorts the light from a background source (such as a more distant galaxy or quasar), resulting in multiple, often distorted, images of the source. The time delay between the arrival of light in these images, the ``time-delay distance," is inversely proportional to the value of \ho, with a smaller dependence on $\Omega_m$ and $\Omega_{\Lambda}$. Time-delay distances are derived by combining detailed modeling of the gravitational potential of the lens with precise measurements of the time delays between the multiple images \cite{refsdal_1964, blandford_narayan_1986}.

In practice, several key steps are involved in this method. First, high-quality imaging data of the lensing system must be obtained, most recently using \hst or ground-based telescopes equipped with adaptive optics. These imaging data are then used to model the mass distribution of the lens, taking into account both luminous and dark matter components. In addition, photometric or spectroscopic monitoring of the background source is conducted to measure the time delays between the arrival of photons in the multiple images. This is a labor-intensive step, requiring observations over several months to years in order to accurately measure the variability and time delays \cite{courbin_2018}. 

Advancements in lens modeling techniques and the quality of data are continually improving \cite{suyu_2017, wong_2020}. Uncertainties in the gravitational lens method arise from the complexity of the lens model, whether the lens is located in a group or cluster, or whether there is mass along the line of sight, as well as due to the assumptions on the cosmological model. An inherent challenge for the method is the `mass-sheet degeneracy’, where an additional underlying mass density (mass sheet) can produce the same deflection angles and magnifications. 
Recently, a joint analysis of six gravitationally lensed quasars with measured time delays \cite{wong_2020} resulted in a value of \ho = 73.3$^{+1.7}_{-1.8}$ km/s/Mpc (a 2.4\% uncertainty), assuming a  flat $\Lambda$CDM cosmology. However, this result is dependent on assumptions about the mass-density radial distribution (e.g., a power-law mass profile) \cite{birrer_treu_2020}. Dropping the assumptions about the mass profile, and  instead using velocity dispersion measurements to  break the mass-sheet degeneracy \cite{birrer_2020}, the precision then drops to 8\%, with \ho = 74.5$^{+5.6}_{-6.1}$ km/s/Mpc. Additional imaging and spectroscopic data for 33 lenses then result in  \ho = 67.4$^{+4.1}_{-3.2}$ km/s/Mpc,  improving the precision to  5\%. Observations and analysis of the multiply lensed SN Refsdal result in values of \ho = 64$^{+11}_{-9}$ \hounits \cite{vega-ferrero_2018} and 64.8$^{+4.4}_{-4.3}$, 66.6$^{+4.1}_{-3.3}$, depending on the model adopted \cite{kelly_2023}. Lensed \sne offer an advantage over lensed quasars due to the increased precision in the time delay measurements, as well as smaller uncertainties in the lens models.

Future improvements to this method will come with larger samples of lenses and measured time delays (to improve the statistical precision), for example, from  the Vera Rubin Observatory, Euclid and the Nancy Grace Roman Observatory, and will require high signal-to-noise kinematic measurements to address the issue of the mass-sheet degeneracy, as well as detailed simulations \cite{ding_2021}.

\subsection{Gravitational Wave Sirens}

Inspiraling neutron star -- neutron star binary systems have offered a new means of measuring \ho that is completely independent of the local distance scale. In analogy with the astrophysical standard candles described earlier, the detection of gravitational waves from these systems provides a `standard siren' that can be used to estimate the luminosity distance of the system out to cosmological distances, without the need for a local (astrophysical distance scale) calibration. The method requires both the detection of gravitational, as well as, electromagnetic radiation (the latter providing the redshift).

 \begin{figure}
    \centering
\includegraphics[width=10.0cm]{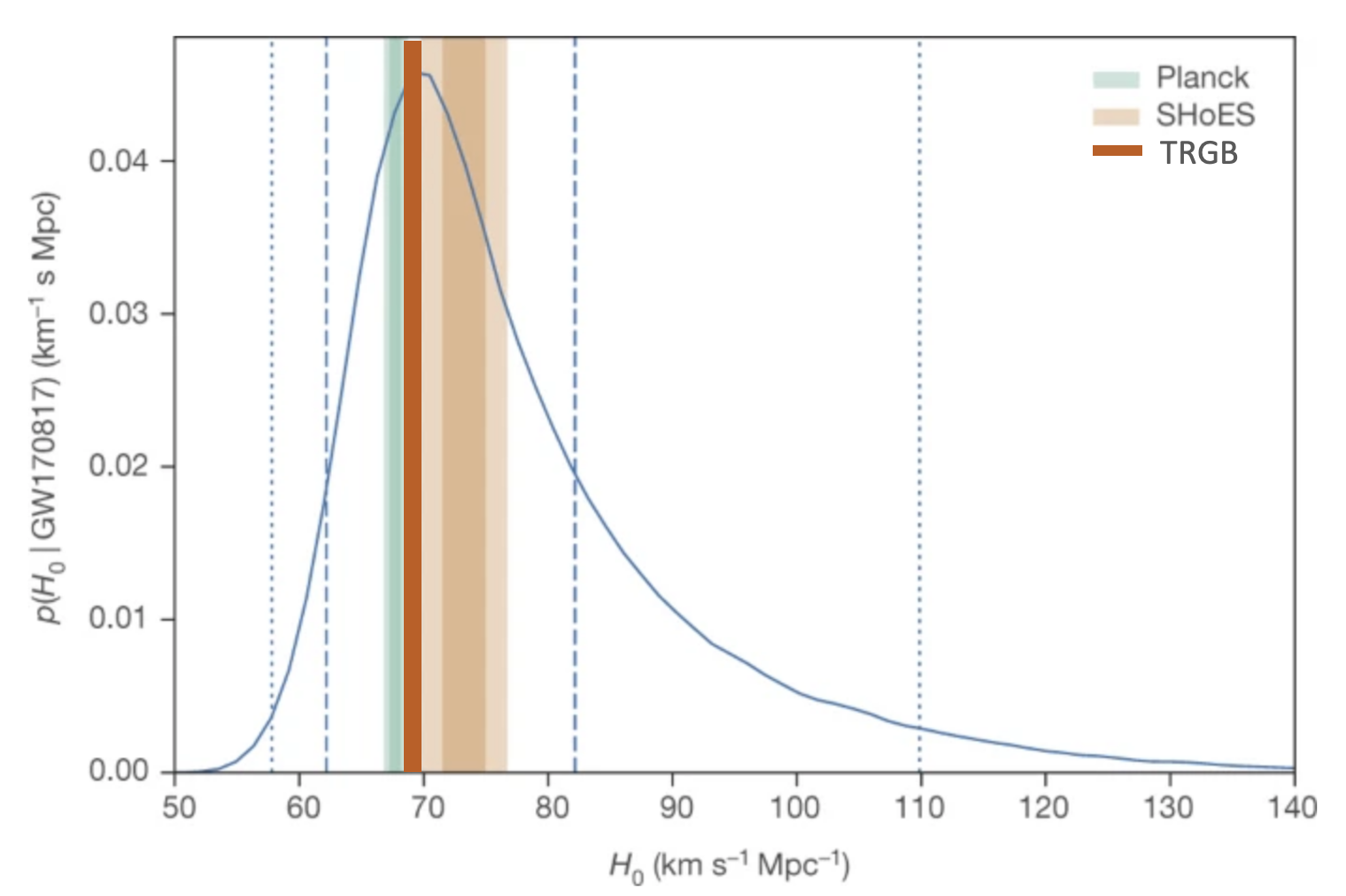}
    \caption{The marginalized posterior density distribution (blue line) for \ho derived from the gravitational wave detection of  GW170817. Constraints from Planck and \shoes are shown in green and orange, respectively. The TRGB value of \ho $=$ 69 \hounits is shown in red. Figure adapted from \cite{abbott_2017}. See text for details.}
\label{fig:GravWave}
\end{figure}

The method was first applied with stunning success to the event GW170817, located in a galaxy at 43 Mpc \cite{abbott_2017}. The authors determined a value of \ho = 70$^{+12}_{-8}$ \hounits (see Figure \ref{fig:GravWave}). A number of factors contribute to the 15\% uncertainty:  detector noise, instrumental calibration uncertainties, uncertainty in the peculiar velocity of the host galaxy, and a geometrical factor dependent upon the covariance of distance with inclination angle. At a distance of 43 Mpc, the peculiar velocity is about 10\% of the measured recessional velocity.

GW170817 was detected with high signal to noise almost immediately after LIGO was turned on in 2017. It led to the expectation that many more sources were likely to follow and that a value of \ho to 2\% accuracy would be possible by 2023 \cite{chen_fishbach_holz_2018} with the detection of 50 events, assuming that redshifts could be measured for each object. Sadly, as of summer, 2023, there have not yet been any comparable events, and an accurate measurement of \ho with this technique will require patience. Ultimately, it will provide a critical independent means of comparison with the local distance scale.

\section{The Hubble Constant and the Impact of the James Webb Space \\ Telescope (JWST)}
\label{sec:jwst}

In this section, we provide an overview, as well as a current status report, of a new CCHP long-term program using \jwst. This program is aimed at reducing the current systematics in the local extragalactic distance scale and the measurement of \ho. Specifically our goals are to: 1) exploit the high resolution of \jwst to understand and reduce the possible effects of crowding and blending of Cepheids previously observed with \hst, 2) improve the corrections for dust, 3) improve the constraints on the metallicity of Cepheids and 4) provide three independent measures (Cepheids, TRGB, JAGB) of the distances to the same galaxies, thereby reducing the overall systematic distance uncertainties.   

The blue sensitivity and high spatial resolution of \hst  made it an ideal facility for the discovery of Cepheid variables. At bluer (optical)  wavelengths, the amplitudes of Cepheid variables are larger than at longer wavelengths due to the greater sensitivity of the surface brightness to temperature, thus facilitating their discovery \cite{freedman_2001}. \hst's high resolution allowed Cepheids to be discovered in galaxies over a larger volume of space than could be accomplished from the ground, most recently  out to distances of $\gtrsim$40 Mpc \cite{riess_2022}.

The superb science performance of \jwst has greatly exceeded early expectations in terms of sensitivity, stability, image quality, as well as spectral range 
 \cite{rigby_2022}. Two key features make \jwst the optimal telescope for addressing the {\it accuracy} of measurements of \ho: its red sensitivity and higher spatial resolution. The extinction is significantly lower: A$_J$ and A$_{[4.4]}$ are smaller by factors of 4 and 20$\times$ respectively, relative to the visual extinction, A$_V$; and factors of 2 and 10$\times$ lower relative to the I-band extinction A$_I$ \cite{cardelli_1989, indebetouw_2005}. \nircam (F115W) imaging from \jwst \cite{rieke_2023} has a sampling resolution four times that of \hst $WFC3$ (F160W),  with a FWHM of 0.04~arcsec on the former telescope and imager, versus 0.151~arcsec on the latter.  In addition, in the near infrared (NIR), the objects that are causing contamination and crowding of the Cepheids  are red giant and bright asymptotic giant branch stars, exacerbating crowding effects in the red, compared to optical wavelengths. {\it Importantly, with ~4 times better resolution than \hst, crowding effects are decreased by more than an order of magnitude in flux using \jwst}.

We have been awarded time in Cycle 1 of \jwst (JWST-GO-1995: P.I. W. L. Freedman; co-I B. F. Madore) to obtain observations of 10 nearby galaxies that are hosts to type \sne, as well as observations of \ngc~4258, a galaxy that provides an absolute calibration through its geometric distance based on H$_2$O megamasers (see Sections \ref{sec:maser_anchor}, \ref{sec:masers}). There are three components to the program: Three independent distances to each galaxy will be measured using Cepheids, the TRGB and \jagb stars, with a particular emphasis on testing for, and decreasing the systematic uncertainties that have often historically plagued distance scale determinations.  The program is designed to deal specifically with known systematic effects in the measurement of distances to nearby galaxies: extinction and reddening by dust, metallicity effects and crowding/blending of stellar images. Simply getting more nearby galaxy distances (decreasing the statistical uncertainties) is insufficient to confirm or refute whether new physics beyond the standard cosmological model is required. At this time, systematic uncertainties are (and  have historically always been) the dominant component of the error budget. Our goal is to decrease the systematic errors to the 2\%-level, for each of the three methods.

The primary sample for the program is a subset of the nearest galaxies that have both reliable \sn photometry and previously-discovered Cepheid variables \cite{freedman_2001,riess_2020}, and for which TRGB and carbon-star distances can now also be measured. All three of these methods individually have high precision and can be independently used to calibrate \sne. The observations are being carried out  in the NIR at $F115W$ (or $J$ band) and mid-infrared  F356W at 3.6$\mu$m  with the \jwst Near-infrared Camera (\nircam), and in parallel at $F115W$-band with the Near-infrared Imager and Slitless Spectrograph (\niriss) \cite{willott_2022}.  Our first observations for \ngc~7250 were carried out with the F444W filter at 4.4$\mu$m, but we have switched to F356W for the rest of the sample, owing to its higher sensitivity and better sampling. However, the F444W filter contains a CO bandhead that is sensitive to metallicity \cite{scowcroft_2016b}, and it is being used to carry out a test for metallicity effects in the galaxies, M101 and \ngc 4258, as discussed further in Section \ref{sec:jwst_cepheids} below.

Our target fields were chosen to maximize inclusion of the largest possible number of known Cepheids in the inner disk, as well as the inclusion of the outer disk to detect carbon stars, and with a rotation angle optimized for the detection of halo red giants. The disk observations are being carried out with \nircam; the outer halo observations with either \nircam or parallel observations with \niriss.  We are carrying out the analysis using two independent software packages, DAOPHOT \cite{stetson_1987} and DOLPHOT \cite{Weisz2023}, in order to provide a quantitative constraint on photometric errors that might arise due to differences in point-spread-function fitting in crowded fields.

In brief, we find (1) The high-resolution \jwst images of \ngc~7250 demonstrate that many of the Cepheids observed with \hst are significantly crowded by nearby neighbors. (2) The scatter in the \jwst NIR Cepheid PL relation is decreased by a factor of two compared to those from \hst.   (3) The TRGB and carbon stars are well-resolved, and with the Cepheid measurements, will allow measurement of three independent distances to each of these galaxies. These new results illustrate the power of \jwst to improve the measurement of extragalactic distances, and specifically, to address remaining systematics in the determination of \ho. 

\begin{figure}
    \centering
\includegraphics[width=8cm]{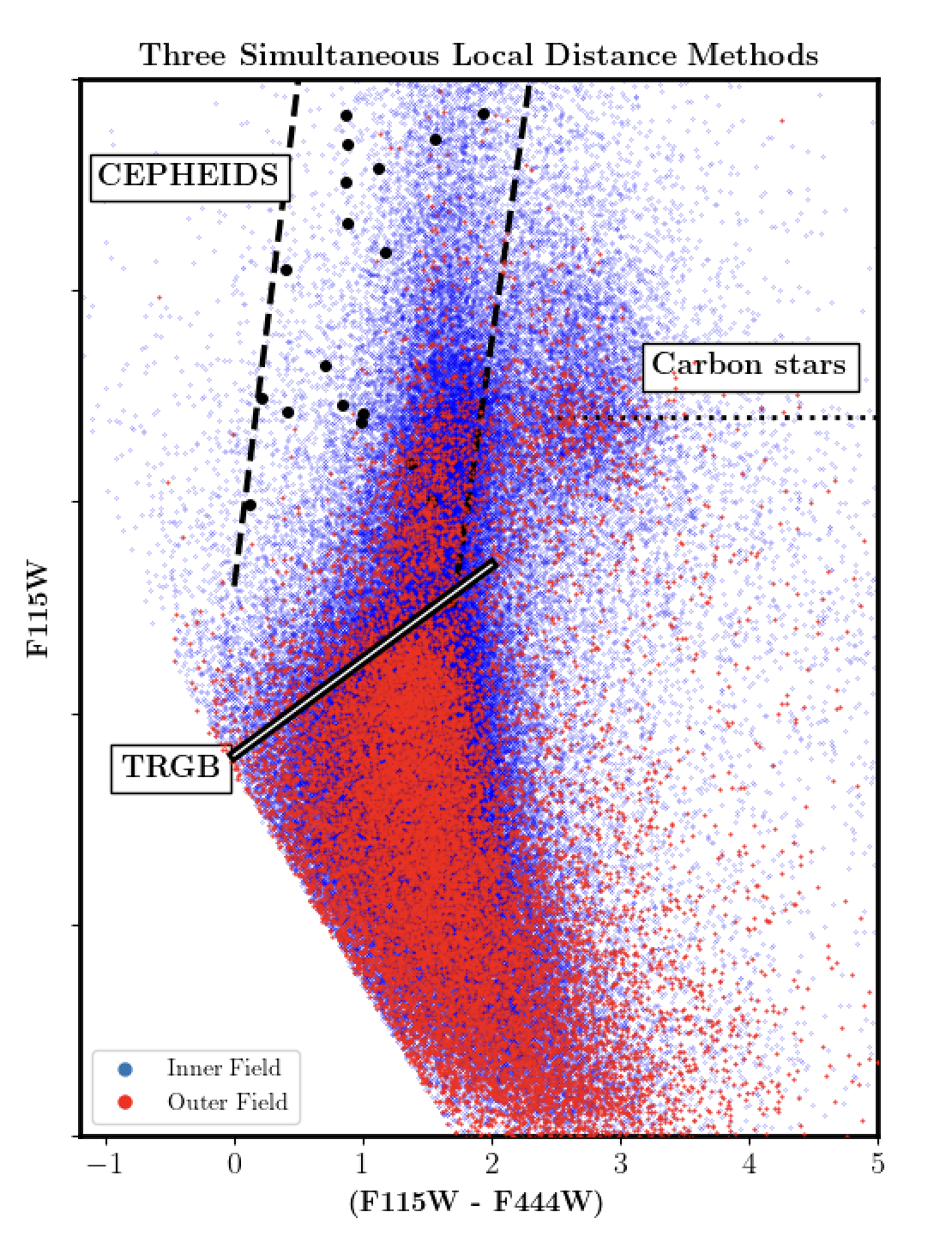}
    \caption{The relative disposition of the three stellar/astrophysical distance indicators, discussed in this review, seen plotted in a \jwst F115W versus (F115-F444W) CMD. Cepheids are the black dots between the two vertical dashed lines, where the latter represent the red and blue limits of the instability strip. JAGB/Carbon stars are  further to the red. Their mean luminosity is marked by the horizontal dotted line. Finally, the TRGB maximum J-band luminosity, as a function of color, is shown by the upward slanting yellow line at the top of the red giant branch at (F115W-F444W)  $\approx$ 1.5 mag. See also Figure \ref{fig:jagb-lee-n7250}.}
\label{fig:CMDComp}
\end{figure}

In Figure \ref{fig:CMDComp} we show a color-magnitude diagram (F115W versus [F115W -- F444W]) for the galaxy \ngc~7250, which shows at a glance, the Cepheid instability strip, the position of the TRGB, the location of the JAGB stars, and the power of this three-in-one program. These three distance scales, all on a common photometric scale, contain valuable quantitative information as to potential systematic differences among the methods. 
The magnitudes are shown on an arbitrary scale, since at this stage of the analysis,  the photometry is blinded. In addition, the current absolute flux calibration for \nircam is only at a level of 5\% (M. Rieke, private communication); however, the desired future  goal is a 1-2\% absolute calibration tied to laboratory-standard measurements \cite{gordon_2022}. 

\subsection{JWST Cepheid Program}
\label{sec:jwst_cepheids}

 \begin{figure}
    \centering
\includegraphics[width=14.0cm]{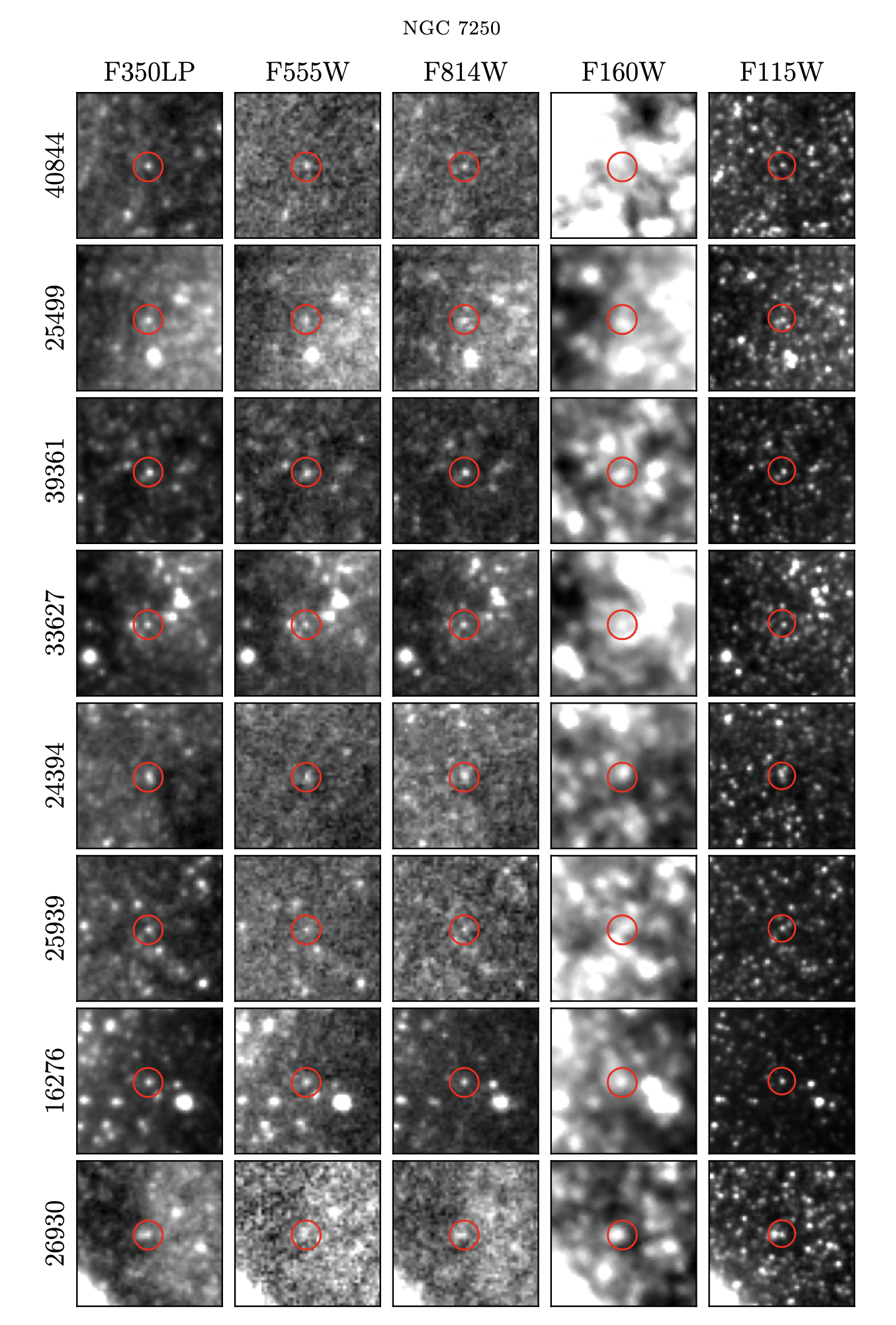}
    \caption{NGC~7250 Cepheids: A sample of cutout images for the light-curve-selected Cepheids in 5 photometric bands. Each
cutout is 2~arcsec on a side. The red circles enclose the location of the Cepheid candidate and are 0.2~arcsec in radius. \jwst J-band images are in the far right column. All other images (all four columns to the left) are from \hst. Adapted from \cite{owens_2023a}.}
\label{fig:N7250}
\end{figure}

 The \jwst Cepheid sample for \ngc~7250 was selected based on a completely new (end to end) re-analysis of the archival \shoes data \cite{owens_2023a}. This archival sample is comprised of 11 epochs of 'white light' (F350LP) photometry, with smaller numbers of (significantly lower signal-to-noise) phase points at three additional wavelengths (three at F555W, two at F814W and six at F160W).  Periods and light curves were measured directly and independently using the F350LP photometry, using templates derived from well-measured Cepheids in the LMC \cite{owens_2023b}. Cepheid variable candidates were selected according to the following criteria: 1) optical colors consistent with known Cepheid variables; 2) optical amplitudes $>$ 0.4 mag; 3) classified according to their light curve quality (requiring a classical `saw-tooth' shape) at F350LP; 4) the light curves and images of the Cepheid candidates were independently inspected by eye by four team members. If there was disagreement about the quality of the candidate, it did not make the final cut; and 5) having no comparably bright nearby companions within the point spread function (PSF) at F350LP, as determined from the higher-resolution F115W data.\footnote{If the summed flux  from resolved sources in the JWST F115W images within 4
NIRCAM pixels of the Cepheid candidate (0.124 arcsec or approximately one \hst WFC3 IR pixel or 0.13 arcsec) was equal to or greater than the measured flux of the star itself, the candidate was
considered to be crowded and not included in the final sample.} These stringent criteria were chosen to reduce the uncertainties due to crowding and low signal to noise. They result in a final sample of 16 uncrowded Cepheids with well-determined light curves. The photometry for all of the Cepheid candidate variables, both before and after final selection, will be made available on github \cite{owens_2023a}.

The new \jwst observations are allowing us to directly assess the degree to which crowding/blending effects have affected the (4$\times$ lower-resolution) \hst photometry, on a star-by-star basis.  In Figure \ref{fig:N7250}, we show multiband cutout images of eight Cepheids in \ngc~7250 at a distance of 20 Mpc. From left to right are images at F350LP, F555W, and F814W (from \hst) and F160W and F115W (from \jwst).  The cutouts are 2 $\times$ 2 arcsec on a side, and  have been scaled as described in the figure caption. These images illustrate the superb resolution and the power of \jwst to improve the measurement of extragalactic distances. The effects of crowding, even in a galaxy as close as 20~Mpc are evident in this comparison. In the \hst data, many of the Cepheid candidates are fainter than their nearby neighbors, rendering background subtraction challenging. \jwst images for the complete sample  of Cepheids  in \ngc~7250 are presented in  \cite{owens_2023a}. 

\begin{figure}
    \centering
\includegraphics[width=10cm]{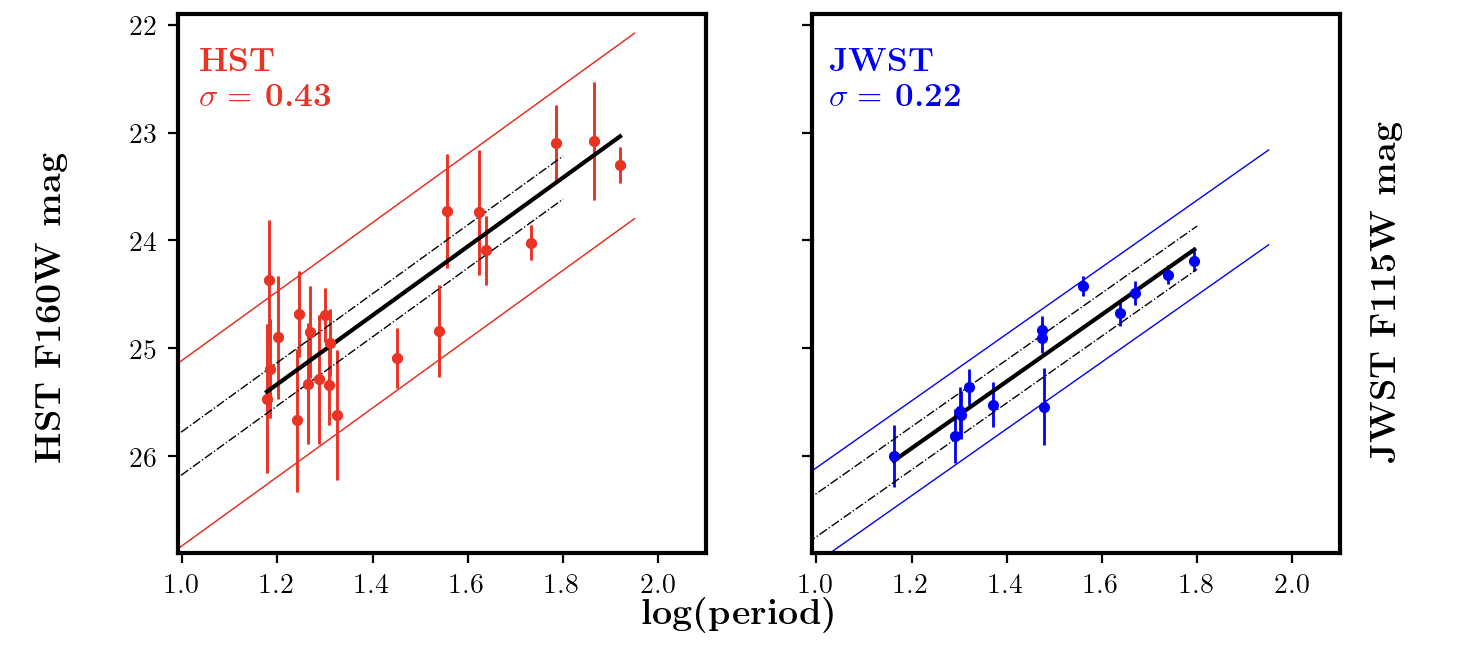}
\caption{NIR Period–luminosity relations for Cepheids in NGC~7250. The left panel is HST F160W (H-band) data from the \shoes collaboration\cite{riess_2022}; the right panel is JWST F115W (J-band) data from the \cchp\cite{owens_2023a}. The scatter about the period–luminosity fit in each filter is labeled in each plot.}
\label{fig:n7250_hst_jwst}
\end{figure}

In Figure \ref{fig:n7250_hst_jwst}, we compare the Leavitt law for Cepheids in \ngc~7250 observed with \hst (left panel) and \jwst (right panel).  The \jwst data are plotted on an arbitrary magnitude scale, as the data are still blinded. The slope is determined from the LMC , and restricted to log P $<$ 1.8, after which the period-luminosity relation shows evidence for non-linearity. The scatter in the \jwst F115W data for \ngc~7250 is a factor of two smaller than the \shoes F160W data, which is all the more remarkable since the F115W data are for a single epoch only. In addition, a two-sigma rejection of candidates in the PL relation has been applied to the \shoes F160W data; no sigma cut has been applied to the \jwst Cepheid candidates based on position in the PL relation. 

When the data are unblinded, and an absolute calibration is established, the \jwst data will allow us to also improve the accuracy of the reddening corrections to the individual galaxies and their Cepheids.  A standard interstellar extinction curve \cite{cardelli_1989,indebetouw_2005}  can be fit to the multi-wavelength $F350, V, I, H$ and  $J$-band apparent distance moduli \cite{freedman_madore_2010, monson_2012}. Finally, the 4.4$\mu$m-band can provide a direct and quantitative measure of the metallicity of the each of the Cepheids. $Spitzer$ 4.5 $\mu$m observations of Cepheids in the Milky Way, the LMC and the SMC revealed a direct correlation between Cepheid metallicity and luminosity \cite{scowcroft_2016b}, a result of a CO bandhead that is present in the 4.4~$\mu$m filter. \jwst observations across the disks of M101 and \ngc~4258 have been scheduled as part of our program. In particular, there is a steep metallicity gradient in M101 \cite{garner_2022}, which will allow a direct test of the metallicity sensitivity at long wavelengths. The uncertainty due to the effects of metallicity remains one of the largest sources of systematic error in the Cepheid distance scale \cite{efstathiou_2020}. 

With improved reddening measurements, a direct measure of the metallicity, and a robust estimate of crowding/blending effects on current samples, we can address three of the largest sources of systematic uncertainty in the local Cepheid distance scale.  
The selection criteria adopted for inclusion in our final sample of Cepheids are deliberately conservative,  with the intention of avoiding systematic effects due to crowding/blending, aiming for quality over quantity. The JWST data  are still blinded, so in the future there will be a significant improvement to the distance measurements.  However, near-IR photometry obtained using \hst in  this galaxy results in a larger scatter due to the lower spatial resolution and the lower signal to noise of  the data. 

It is important to keep in mind that crowding effects will become more severe with increasing distance. We note that 60\% of the Riess et al. (2022)  sample of galaxies in which Cepheids have been discovered lie at greater distances than \ngc 7250 at 20 Mpc, and that 25\% of the sample lies beyond 40 Mpc. At a distance of 40 Mpc, four times the area will be contained within a given pixel. For the most distant \shoes galaxy at 80 Mpc, 16 times the area will be covered. As the need for percent-level accuracy has grown in importance, and given the level of crowding that we have seen for Cepheids in a galaxy at a distance of 20 Mpc, it remains important to demonstrate that crowding effects do not produce a systematic bias in the photometry and hence, the distance measurements for these more distant galaxies observed with \hst. 

\subsection{JWST Tip of the Red Giant Branch Program}
\label{sec:jwst_trgb}

As noted in Section \ref{sec:trgb}, the TRGB provides one of the most precise and accurate means of measuring distances in the local universe \cite{freedman_2019}. The observed color-magnitude diagrams (CMDs) of the halos of nearby galaxies reveal a sharp discontinuity in the magnitude distribution of red giant branch stars at a well-determined luminosity,  which corresponds to the location of the core helium-flash.

Measuring the TRGB in the near-IR has a number of advantages over the optical: 1) The extinction is significantly lower 2) TRGB stars are brighter in the NIR (M$_J$ = --5.1 mag \cite{madore_2018}) than in the optical (M$_I$ = --4.05 mag \cite{freedman_2020}), making them comparable to that of Cepheids with periods of 10 days (M$_J$ (10-day Cepheid) = --5.3 mag \cite{persson_2004}). The slopes of the RGB as a function of wavelength are well-defined \cite{madore_freedman_2020, madore_2018, hoyt_2018, durbin_2020}. 
3) The peak luminosity of the giants occurs at NIR wavelengths.  The disadvantage of the near-IR is that because the magnitude of the TRGB is no longer flat, as it is in the I-band, it necessitates more accurate measurements in a second filter to measure the slope of the RGB.

\begin{figure}
    \centering
\includegraphics[width=\columnwidth]{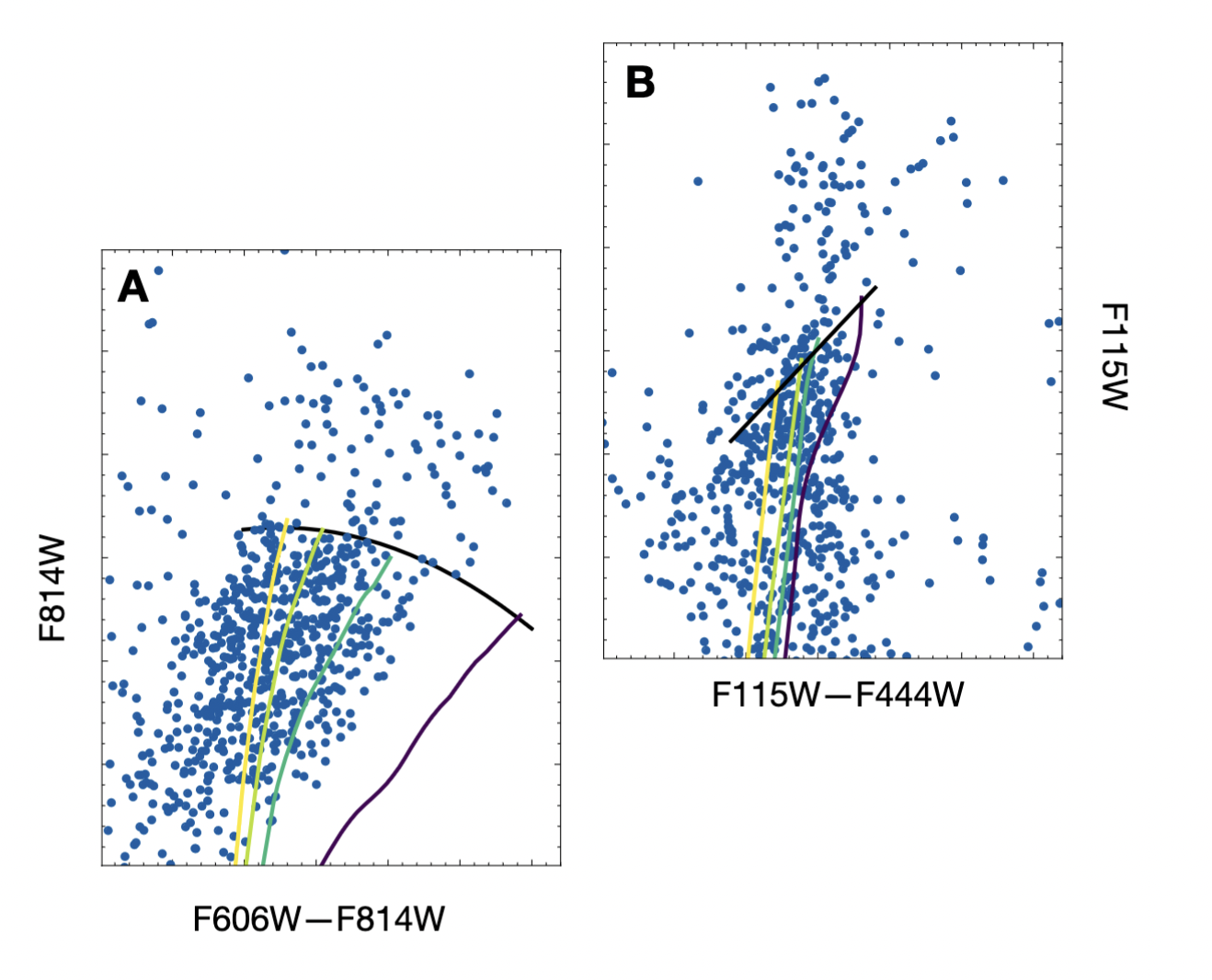}
    \caption{Optical \hst (left) and near-infrared \jwst (right) CMDs for stars located in the stellar halo of \ngc~4536. The CMDs are aligned on their vertical axis to demonstrate the increasing brightness of RGB stars when observed in the infrared. The \hst images represent a  total of 14,000s in telescope exposure time, while the \jwst images represent just 2,800s of exposure time. The known, shallow color dependence of the optical TRGB is overplotted on the left, while the theoretically-predicted slope of the infrared TRGB is overplotted on the right, both as black curves. In both CMDs, 10 Gyr theoretical stellar evolutionary tracks are shown and colored from light yellow to dark purple for metallicities Z = {0.002, 0.004, 0.008, and Z$_\odot$}. The isochrones are shifted to terminate at the observed level of the TRGB.}
\label{fig:hoyt_n4536}
\end{figure}

As part of our \jwst \cchp program, the TRGB has been measured in  \ngc~4536, a galaxy located in the constellation Virgo, about 10 degrees south of the center of the Virgo Cluster.  In Figure \ref{fig:hoyt_n4536} we show an $F814W$ versus $[F606W$ - $F814W]$ color magnitude diagram (CMD) [left panel] and  $F115W$ versus $[F115W$ - $F444W]$ CMD [right panel]  \cite{hoyt_2023b} for \ngc~4536. The downward-arching black curve in the middle of the left panel  illustrates the shallow color dependence of the TRGB at optical magnitudes \cite{jang_lee_2017a}. The theoretically predicted slope of the infrared TRGB is also shown in black in the right panel. Also plotted are stellar evolutionary curves, as described in the figure caption. All magnitudes shown are on an arbitrary scale, but the two panels are aligned, illustrating the brighter magnitudes of the TRGB in the near-infrared relative to the optical. Once again, \ngc~4258 will ultimately provide a geometric zero-point calibration. See \cite{hoyt_2023b} for details of the analysis of these data. 

\subsection{JWST Resolved Carbon-Rich AGB Stars Program}
\label{sec:jwst_jagb}

\begin{figure}
    \centering
\includegraphics[width=\columnwidth]{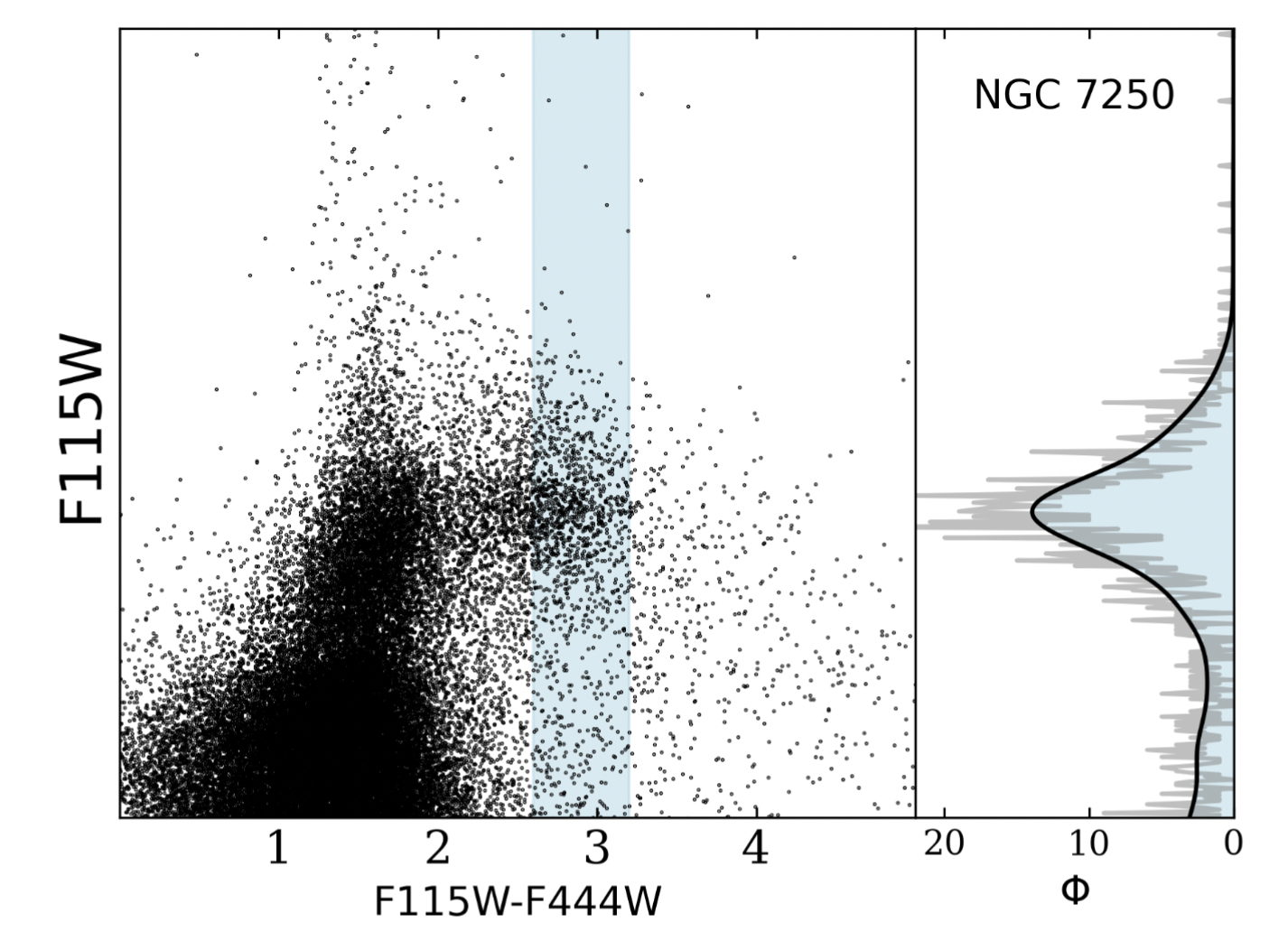}
    \caption{$F115W$ versus [$F115W$- $F444W$] color-magnitude diagram for the outer region of the galaxy \ngc~7250 (left panel) from \cite{lee_2023b}. The JAGB stars were measured to be within the light blue shaded region. In the right-hand panel, the GLOESS-smoothed luminosity functions for the JAGB stars is shown in light blue, and the 0.01~mag binned luminosity functions are shown in grey.  Within a window of 1.50~mag wide centered on the mode, the scatter of the JAGB stars is $\sigma=0.32$~mag. }
\label{fig:jagb-lee-n7250}
\end{figure}

In Figure \ref{fig:jagb-lee-n7250}
we show an $F115W$ versus [$F115W - F444W$] CMD for the outer disk of the galaxy \ngc~7250, which illustrates immediately the feasibility of using \jwst and this method for distance determination. The carbon stars are located to the red of the TRGB, about one magnitude brighter than the tip, and exhibit a nearly-constant luminosity with a dispersion of only $\pm$0.3 mag. These single-phase observations have only a slightly larger scatter than the intrinsic (time-averaged) scatter observed in the LMC \cite{weinberg_nikolaev_2001}. Details of the analysis, as well as for the galaxies \ngc~4536 and \ngc~3972  are presented in \cite{lee_2023b}.

\section{Is There a Crisis in Cosmology?}
\label{sec:crisis}

\begin{figure}
    \centering
\includegraphics[width=\columnwidth]{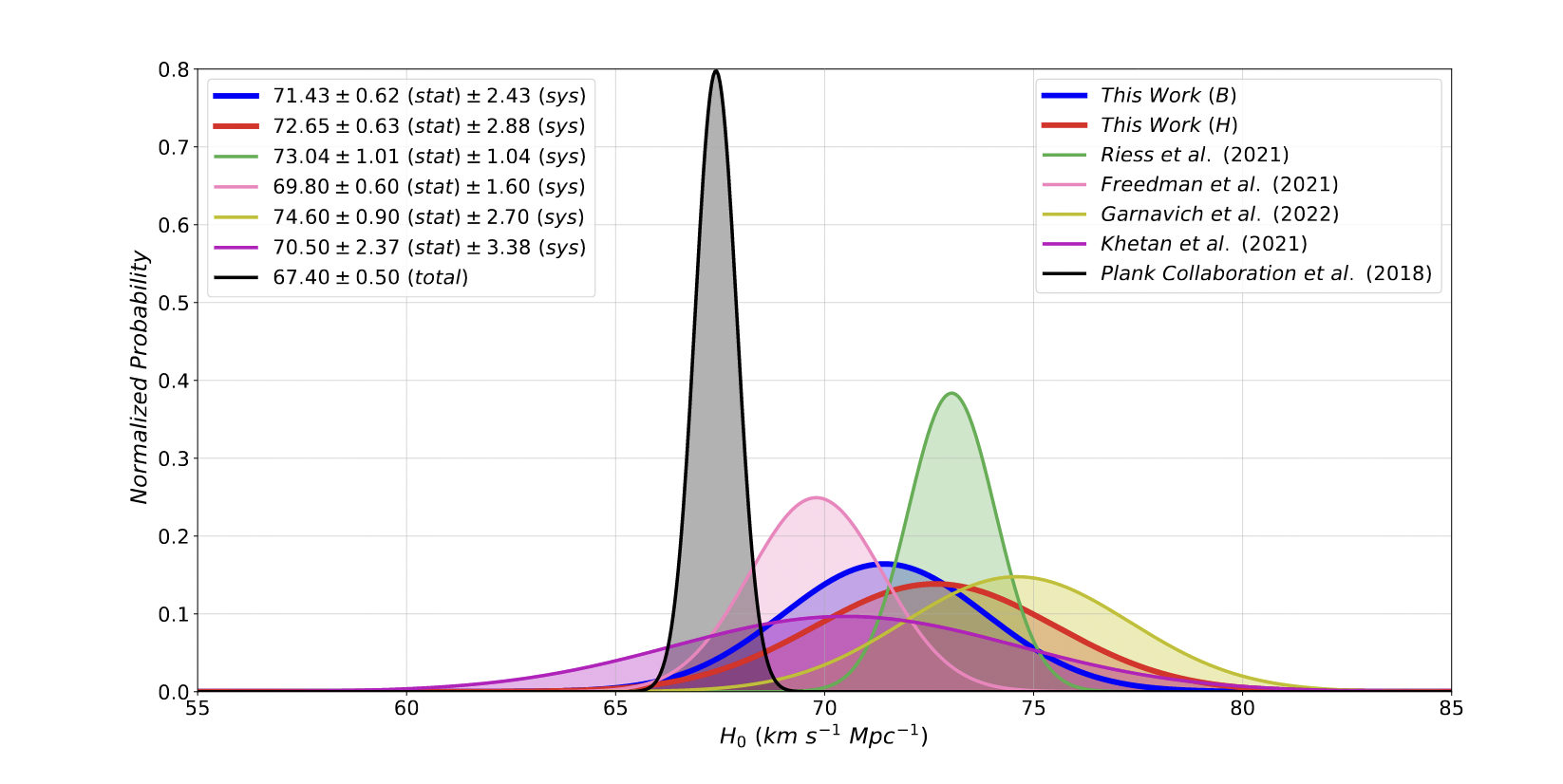}
    \caption{Probability distributions for \ho for calibrations based on Cepheids \cite{riess_2021}, the TRGB \cite{freedman_2021} SBF from \cite{uddin_2023}, compared to recent published values from the literature. The Planck Collaboration value from the CMB \cite{planck_2018} shown in grey.}
\label{fig:uddin_2023_ho}
\end{figure}

Time will tell if cosmology is facing a crisis. It still remains at a crossroads\cite{freedman_2017}. The precision and accuracy with which extragalactic distances can be measured continue to improve, and many new facilities/programs are now either ongoing or will be online in the near future, which will lead to the continued refinement of the distance scale and to the measurement of \ho. In Figure \ref{fig:uddin_2023_ho} we show a comparison of recently published values of \ho from \cite{uddin_2023}. To date, none of the local measurements of \ho reach the $<$1\% precision of the Planck result that is inferred from CMB measurements. 

At this juncture, and given the still outstanding issues that need to be unambiguously addressed in order to allow a 1\% measurement (e.g., small numbers of anchors, crowding effects, consistency across observing wavebands, metallicity effects), it is reasonable to keep an open mind as to the ultimate resolution of this latest crisis. 

The current outstanding question essentially now revolves around `{\it the uncertainty in the uncertainty}'; i.e., have we yet reached a level of precision and accuracy in the local distance scale that can test the CMB model, which itself is quoted to have a precision exceeding 1\%. 5$\sigma$ in experimental physics is the gold standard.  How robust is the currently claimed astronomical 5$\sigma$ result? If the result is secure at the 5-6$\sigma$ level, then in principle, the question is settled, and no more work need be done. It is perhaps illustrative, however, to consider that if the uncertainty in \ho were to have been underestimated by only a factor of 1.5 and \ho = 72.0 $\pm$ 1.5 \hounits, then the tension with the Planck results drops from 5$\sigma$ to less than 3$\sigma$. Similarly if \ho = 72.0 $\pm$ 2.0 \hounits, the tension drops to 2$\sigma$.

\section*{Summary}
\label{sec:summary}

The advancement in measuring the distances to galaxies over the past twenty-five years has been nothing short of remarkable. Just two decades ago, achieving accuracies within a few percent for the extragalactic distance scale was virtually unthinkable. This  progress can be attributed to better detectors, increased wavelength coverage, innovative new, independent methods for measuring distances, and access to space, all of which have made it possible to address systematic effects including reddening/extinction from dust, metallicity, and crowding.

The launch of \jwst has opened a new chapter in the measurement of extragalactic distances and \ho. The superb resolution and unequalled sensitivity at near-infrared wavelengths is already demonstrated in the first data from the nearby galaxies, \ngc~7250,  \ngc 3972  and \ngc~4536,  at distances between $\sim$15-20 Mpc.  These early data clearly demonstrate the promise of \jwst for improving the measurement of extragalactic distances and the local, directly measured value of \ho. Our program has been optimized to observe Cepheids in the spiral arms of the inner disks of galaxies, JAGB stars in the extended disks, and TRGB stars in the outer halos of galaxies. All ten of the program galaxies are \sn hosts; an eleventh galaxy, \ngc~4258,  will provide an absolute distance calibration through the geometric measurement of its distance based on H$_2$O megamasers. 

For the first time, we have \jwst data for Cepheids where stars located within  one PSF radius, that were discovered on \hst frames, can be directly identified. Limiting the sample of Cepheids to exclude the variables with nearby neighbors, results in a distance modulus that is +0.45 mag farther away (in the sense that its contribution would result in a lower value of \ho).  Future data will reveal whether this is indicative of a systematic effect to be found in the larger sample. 

While it has become a  common refrain  in the literature that systematic effects can no longer be considered as relevant for the \ho tension, with differences this large in a galaxy at only 20 Mpc distance, there are reasons to keep open to the possibility that ``unknown unknowns", or perhaps ``known unknowns", could still be significant.\footnote{With acknowledgement to Donald Rumsfeld who said ``There are known knowns. These are things we know that we know. There are known unknowns. That is to say, there are things that we know we don't know. But there are also unknown unknowns. There are things we don't know we don't know.”} 

How then could we imagine that a difference, with a quoted significance of 5-6$\sigma$, is still possible to reconcile? As we have seen, H-band (F160W) photometry  obtained with \hst in crowded fields appears to be highly challenging, with uncertainties far exceeding the currently quoted uncertainties on the Cepheid calibration of the local distance scale. Another example of the challenge  comes from an comparison of the F160W photometry published by \shoes between 2016 and 2022. We note that even internally within the \shoes program, large differences for the same galaxies are seen in an earlier  analysis of the same data following a later analysis, even by the same group. Moreover, and of more concern as the goal is to achieve 1\% level accuracies, these differences  are systematic in nature. A simple comparison of the F160W distance moduli tabulated in R16 and R22  reveals an overall difference of   $-$0.123 with an $rms$ dispersion of 0.085 mag\footnote{Determined by computing a mean difference for all Cepheid fluxes in each \sn host galaxy and taking the mean of those values. These statistics were determined for 874 Cepheids in common between R16 and R22 observed across 21 host galaxies; Hoyt, private communication.}, with five galaxies having large differences ranging from -0.2 to -0.7 mag.  (Using the median instead, the overall  difference between R16 and R22 is $-0.096 \pm 0.068$ mag.) The above-quoted difference of -0.123 mag, on average, corresponds to a 6\% shift in \ho. Put another way, if the newer, more reliable F160W magnitude measurements (according to R22) had been available in 2016, it would have resulted in a value of \ho = 68.97  \hounits (and a corresponding tension of only 0.9~sigma compared to the Planck value of 67.4 $\pm$ 0.5 \hounits). Given that the reduction of the \hst near-IR photometry is still in a state of  flux, as found in our independent reduction (details in \cite{owens_2023a}) and the \shoes own independent analysis, it surely must signal that better near-IR data, at least, are essential to resolving both the `local' and the CMB-versus-Cepheid \ho tensions.

Over the next year, we will continue to obtain data and complete the analysis of our \jwst \cchp  program sample of ten galaxies, all with distances $\lesssim$ 20 Mpc,  chosen to be close enough to minimize potential crowding effects.  Ultimately, we will calibrate \sne based on three independent techniques --  Cepheids, the TRGB and JAGB/carbon stars -- and determine a value of \ho with significantly reduced systematic uncertainties (including reddening by dust, differences in chemical composition, and crowding/blending of images).    These data will allow us to provide an answer to one of the most important problems in cosmology today -- Is there new fundamental physics required beyond standard $\Lambda$CDM?

%\subsection{Comparison of SHoES and CCHP distances}

\section{Appendix: Intercomparison of the Cepheid, TRGB and JAGB \\ Methods and Sensitivities}

\subsection{General Remarks}

Having introduced each of the primary extragalactic stellar distance indicators, Cepheids in Section \ref{sec:cepheids}, TRGB stars in Section \ref{sec:trgb} and JAGB Stars in Section \ref{sec:jagb}, we now inter-compare the pros and cons, strengths and weaknesses of each of the three methods to various known, as well as further potential sources of systematic and statistical errors. Some of these errors are a strong function of distance, others are functions of color/metallicity, surface brightness and star formation history, with each of them changing in amplitude, and possibly in sign, as a function of wavelength/bandpass. Theory can act as a guide or even offer some understanding of the distance indicators and the sense of changes resulting from certain underlying physical parameters being varied; but none of the three types of stars being discussed here have calibrations that are derived strictly from ``first principles''.\footnote{The only one that come close is the TRGB.}  

The impact of some of the errors on these distance indicators can be controlled by optimizing the observations ahead of time (e.g., moving to the infrared to minimize reddening, or only observing in the halo, for TRGB stars), others can be dealt with after the fact (e.g., using a priori knowledge of light curve shapes and amplitudes as a function of wavelength in deriving mean magnitudes); still others (such as the total number of Cepheids to be found in a given galaxy) are simply facts to be dealt with, with those limited numbers quantified. Accurate photometry is essential in all cases.  Finally, quantifying the sky backgrounds, and undertaking corrections due to contaminating stars both remain challenging for stellar photometry being undertaken in crowded fields. 

The discussion below is intended to illustrate current or potential additional areas of weakness in all of the methods that have not yet been addressed, but may be important in trying to reach a goal of 1\% accuracy. However, a single path will not be sufficient. As was done for the \hst Key Project, the goal must be to find a solution where many paths of high accuracy converge to a mutually consistent answer.

The three distance indicators under discussion here, are drawn from three distinct and identifiable stellar populations, differing by their typical masses, their evolutionary stages, their variability (or not), their intrinsic spatial distributions within their host galaxies and last, but not least, their individual metallicities (in the interior and in their atmospheres, which may or may not be the same).

\subsection{Photometry}

Photometric accuracy is a key ingredient to the ultimate determination of the local distance scale; however, obtaining accurate photometry in the target fields in galaxies sufficiently distant to host \sne is non-trivial. Several crowded-field software packages are available (e.g., DAOPHOT/ALLFRAME \cite{stetson_1987, stetson_1994}, DoPHOT \cite{schechter_1993, saha_2000}, DOLPLOT \cite{dolphot_software}). Bounds on the systematic uncertainties can be obtained by using more than a single analysis package, as for the case of the Key Project \cite{freedman_2001} with DAOPHOT/ALLFRAME and DoPHOT. A careful study of the galaxy \ngc 3370 using DAOPHOT and DOLPHOT \cite{jang_2023} reveals that the statistical errors returned by the photometry codes are 25-50 \% smaller than the errors measured from artificial star tests. While statistical uncertainties can be overcome by having larger samples of stars, the same is not the case for systematic errors. The latter are magnitude dependent and become larger at the faint end, at the level of $\sim$0.1 mag (5\% in flux, and ultimately, distance). These kinds of (rather significant) systematic uncertainties are often not included in the final error budgets for \ho. 

For a broad-based simulation of the effects of photometric errors, crowding and smoothing of the data for measurements of the TRGB see \cite{MF_23}.
As discussed in the literature \cite{monelli_2010, jang_2023} these differences amongst different software packages likely result from different choices of input parameters, including 
sky annuli, fitting radii, and PSF models. Ascertaining the `correct' solution and completely eliminating the systematic effects may not yet be feasible, but the differences should be reflected in the overall uncertainty.

\subsection{Crowding}

With increasing distance even an entire galaxy of stars will dissolve into a single, spatially-unresolved point of light. Aided by larger and larger aperture telescopes many of the galaxies, diffuse to the unaided eye, can be resolved into individual ``brightest'' stars, while the more numerous and fainter stars are reduced to the status of ``surface brightness''. Crowding is inevitable. Even from space, crowding limits our ability to detect and/or measure stars projected onto the main bodies of individual galaxies.   

To first order, at a fixed distance, crowding is directly correlated with the local surface brightness. Virtually all of the continuum radiation, making up the surface brightness immediately surrounding any given stellar distance indicator, is itself ultimately due to individual stars (resolved or not). Those stars that are bright enough to be detected individually, transition to being called ``sources of crowding''. It is their spatially stochastic appearance, around and under the point spread function (PSF) of the stellar distance indicators, that is particularly vexing. For any given stellar distance indicator we can be informed as to the likelihood of it being crowded, but what exactly is hidden under its PSF can only be determined with any certainty using higher resolution imaging data to actually ``look underneath", at which point the issue of crowding becomes moot ... for that wavelength. 

When attempting to deal quantitatively with crowding, at least four ways forward suggest themselves: (a) Running artificial star tests on the available imaging data, (b) Obtaining higher resolution imaging with the same telescope, but at shorter wavelengths, (c) Obtaining higher resolution imaging with a larger telescope operating at the same wavelength, or (c) Moving the application to lower surface brightness regions further out in the galaxy (e.g., \cite{Maj_2014}). In some cases, Cepheids for instance, moving out of the star-forming region of the galaxy may not be an option.) 

\subsection{Mass}

Differences in the instantaneous masses of stars, indicative of the three methods being discussed here, manifest themselves differently for each type.

The masses of Cepheids (tentatively gauged by their main sequence progenitors) are tightly tied to the periods of these variable stars. After leaving the main sequence the high-mass O and B-type stars that are en route to becoming Cepheids, evolve in the first instance at approximately constant luminosity across the Hertzsprung-Russell (HR) diagram, and into the Cepheid instability strip. Higher-mass Cepheids have lower mean densities and therefore longer periods. But the mass mapping is not unique. First-crossing Cepheids quickly traverse the instability strip and their variability ceases. They then become red supergiants, increase their luminosities and loop back to the blue re-entering the instability strip. At this point the unique tagging of mass to luminosity is broken, and then becomes triply ambiguous when the Cepheid pivots back to the red, once again increasing its luminosity as it does so. Each one of these crossings would be characterized by its own period-luminosity relation. While theory tells us that most of the time is spent in the second crossing, the presence of first and third-crossing Cepheids introduce irreducible scatter into the composite PL relation. If measuring metallicities for individual extragalactic Cepheids is unlikely, then the possibility of measuring individual masses of those same stars at tens of megaparsecs is less so.

Possible mass loss in the red supergiant phases of a Cepheid's life cycle potentially makes the situation more complicated. If that mass loss is a function of metallicity (or, even worse, if it is stochastic) then standardization is problematic without individual metallicities being measured for individual stars. While these issues may not have been a problem for measuring distances at the 10\% level, it remains to be demonstrated that they are not a very real problem in an era where 1\% is the desired goal.

For TRGB stars the situation with regard to masses of the stars populating the red giant branch is  more straightforward. As noted in Section \ref{sec:trgb} the evolution of stars up the red giant branch is completely controlled  by the (detached) evolution of the degenerate helium core as it grows in mass from the ``ash'' raining down upon it from hydrogen burning in the shell directly above it. As more helium falls onto the core, the core contracts, the temperature rises and the energy output of the shell accelerates. The amount of mass in the envelope above the core turns out to be largely irrelevant to the evolution and/or instantaneous properties of the core and its shell; the envelope is simply a vertically inflated source of fuel feeding the shell. At a well defined  mass the core ignites the triple-alpha (helium burning) process.  It does so at a fixed temperature, defined by well-established laboratory nuclear physics, at a fixed radius and at a fixed luminosity of the shell, all of which are independent of the 
{\it total} mass of the red giant star (and {\it inter alia}, independent of any mass loss that may or may not occur or not during that ascent.) Bolometrically, TRGB stars are as close to being standard candles as one can hope for \cite{Bella_2001}. 

JAGB stars are excellent distance indicators because of the mass-sensitive processes that are integral in down-selecting the far more numerous (and much more widely-spread-in-luminosity) hot/``blue" oxygen-rich AGB stars, thereby producing the much cooler/redder carbon-rich AGB stars within which we find the JAGB population of distance indicators (see Section \ref{sec:jagb}).

\subsection{Evolutionary Status}

The advanced evolutionary phases of Cepheids, as they cross the instability strip multiple times, are a strong function of their metallicities. This is especially true of their second and third crossings that are the result of ``blue loops" following their evolution through the red supergiant phase outside and to the red of the instability strip. As illustrated in Figure 13, moving from the bottom, low-metallicity, panel (Z = 0.001) to the top, high-metallicity,  panel (Z = 0.02), the lowest mass Cepheids especially have their blue loops shortened systematically as a function of increasing metallicity. At the highest metallicities (top panel) the blue loops are so shortened that the instability strip is totally devoid of Cepheids. Incomplete filling of the strip as a function of period could give rise to changes in the apparent width and the measured slope of the resulting PL relation, as a function of metallicity. This would be in excess of any wavelength-dependent (atmospheric) metallicity effects on the  colors and luminosities of Cepheids penetrating the strip.

As already discussed above, both the TRGB and JAGB stars are evolutionarily selected by physical processes that are strictly controlled by mass. 
\begin{figure}
\centering
\includegraphics[width=8cm]{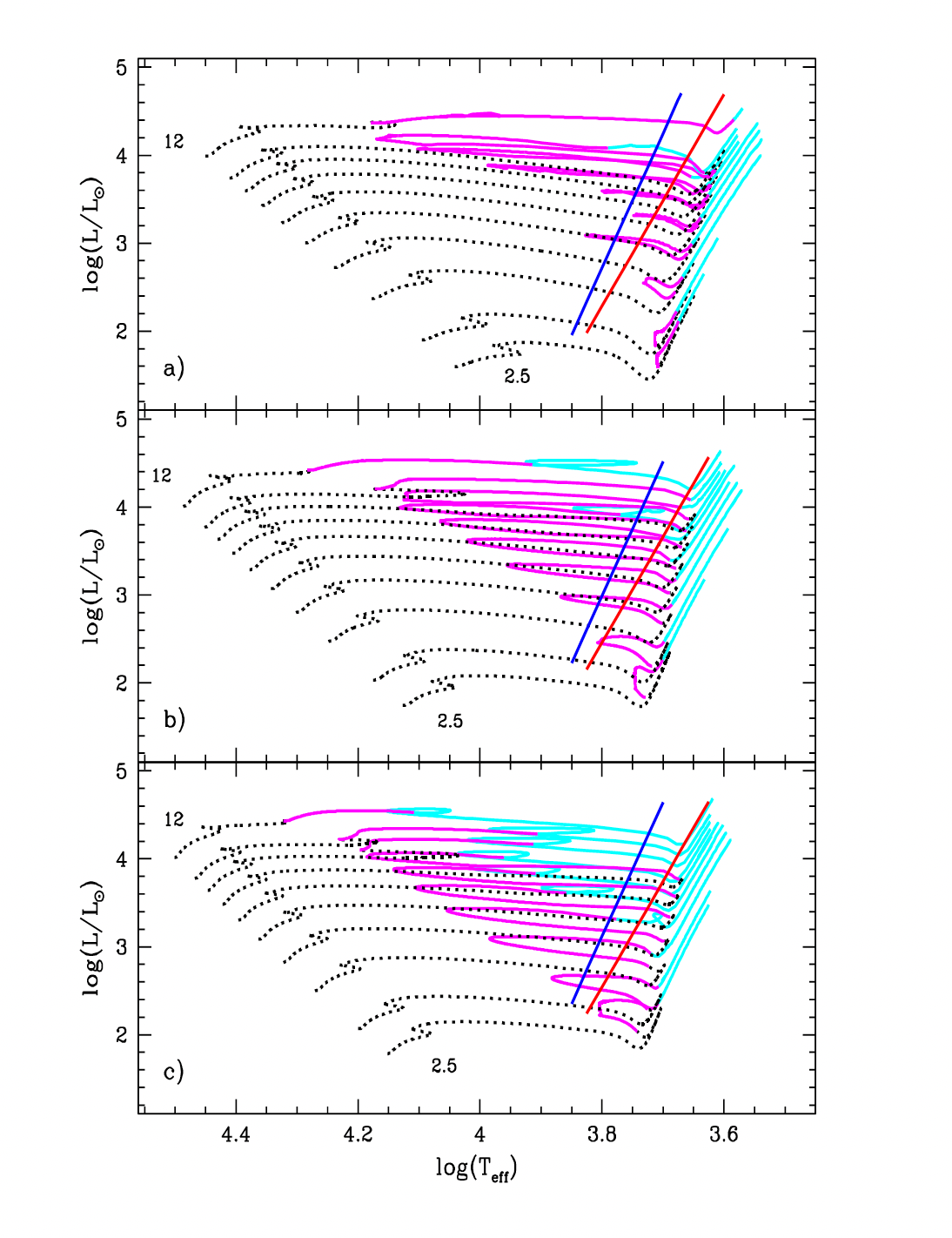}
\caption{Three evolutionary HR diagrams for stars having masses ranging from 12 to 2.5 solar
masses. The Cepheid instability strips are shown by vertically sloping blue and red lines. Increasing
the metallicity (bottom to top) from Z = 0.001 to 0.01 and finally 0.02 illustrates the progressive
depopulation of the short-period region of the Cepheid PL relation, with higher metallicity stars
pulling to the red and systematically failing to make it back into the instability strip in their blue loop and attempted second
``crossing”. Adapted by Bono from \cite{Somma_2021}.
}
\label{fig:HR}
\end{figure}

\subsection{Spatial Distribution}

As already noted, Cepheids are evolving, high-mass stars whose progenitors are young, hot O and B type stars. As such, Cepheids are still physically close to sites of star formation which include their progenitors, as well as the residual gas and dust from which they are formed. These regions are also of higher-than-average surface brightness both because of the general star formation activity at blue wavelengths, but also in the red where density waves, when they are actively involved in spiral structure, will collectively concentrate the low-mass red stars as well. Owing simply to their young ages, Cepheids are only to be found in gas-rich, dusty, high-surface-density regions of spiral and irregular galaxies, prone to large and variable amounts of total line-of-sight extinction and greater-than-average amounts of crowding and confusion.

TRGB stars are found at the opposite extremes of each of the above  situations that Cepheids find themselves in. TRGB stars are old, Population II stars that are denizens of the halos as well as the inner bulges of galaxies. An advantage is that they can always be found far from the disks of their parent galaxies where there is little or no gas or dust to dim/redden them. With the exception of a low level contribution of AGB stars, the dominant population of resolved stars in the halo is the TRGB population itself. If crowding is to occur it will statistically be RGB stars crowding other RGB stars. An easy calculation or a quick look at any
given frame will show just how densely packed the TRGB stars are, and whether a more sparsely populated region should be selected or not. In any case, there are good reasons  not to use TRGB stars as distance indicators if there is evidence (in the color-magnitude diagrams themselves) for younger populations of blue or red supergiants,
or large numbers of intermediate-age AGB stars in the field. Such fields will have dust, and they will be more susceptible to crowding by their interloper populations. Rather than trying to compensate for working in a compromised field  in terms of potential systematics, it is best to address all of the associated problems by simply moving farther out into the halo.

JAGB stars are an interesting population of stars that are very numerous, old enough to be smoothly distributed in space (e.g, see Figure 9   \cite{Rowe_2005}), but still associated with the disk and its gas and dust. However, they do extend well beyond the most dusty regions, well out into the extended disk (possibly even defining it). Moreover, since the luminosities of JAGB stars are specifically defined to be measured in the J band (at 1.2 microns), the effects of dust are diminished with respect to the optical.

\subsection{Metallicity}

Each of our three stellar distance indicators independently span a range of (largely non-overlapping) metallicities. Cepheids are high-mass, high-metallicity, young  Population I stars. JAGB stars are intermediate-mass, intermediate-age, and intermediate (interior) metallicity stars whose polluted surface abundances bear no resemblance to their immediate progenitors, nor to their main sequence star progenitor metallicities.  TRGB stars are old, low-mass, Population~II stars that cover a range of low metallicities.

Empirical studies of the effect of atmospheric metallicity on the magnitudes and colors of Cepheids are still actively debated (\cite{owens_2022} \cite{Breuval_2022} and references therein), not only in the optical, but now in the near and mid-infrared  \cite{breuval_2021} where expectations were that line blanketing effect at least would be minimal.

Theoretical studies of the effects of interior metallicity on the evolutionary tracks of Cepheids criss-crossing the instability strip indicate that the second crossing, and especially the degree to which it penetrates the allowed region of Cepheid variability before the star turns and evolves back out of the strip to the red, is indeed a function of metallicity. 
As illustrated in Figure \ref{fig:HR}, moving from the bottom panel (Z = 0.001) to the top panel (Z = 0.02), the lowest mass Cepheids especially have their blue loops shortened systematically as a function of increasing metallicity.
Incomplete filling of the instability strip as a function of mass (as predicted to be the case) would manifest itself as being a function of period, and would  of course, change the slope and zero point of that population's PL relation.

JAGB stars, by their very nature, have atmospheres that are totally dominated by carbon that has recently been formed and convected to the surface. Whether that carbon gets to the surface or not is determined by envelope physics that is controlled by the mass of the star. Any direct knowledge of the interior metallicity of a JAGB/carbon star is completely masked by the overwhelming presence of recently introduced carbon in the atmosphere.

A fine introduction to the mapping of theory to observations of TRGB stars is found in \cite{Bella_2001}. Using the linking equations found therein it can be shown that from $-2.0 < \rm{[Fe/H]} < -1.0$ dex it follows that $1.39 < (V-I)_o < 2.12$ yielding $-4.04 > M_I > -4.01 $~mag. So to within $\pm$0.015~mag $M_I$ is a standard candle, independent of color or metallicity over range cited above. With that in mind, \cite{Tik_21} have made the case for applying the TRGB method using only I-band luminosity functions without a second filter (or a CMD) being required.

\subsection{Binarity}

The incidence of Cepheids having companions is thought to be at least 30\% 
 \cite{Madore_1977, Evans} and because of evolutionary timescale differences most of the companions are probably lower mass, bluer and certainly of lower luminosity than the parent Cepheid. How that incidence of binarity changes from galaxy to galaxy or within galaxies (as a function of metallicity, say) is unknown. The influence of companions on observed PL relations have been recently discussed in \cite{Karc_2022, Karc_2023}. It is important to treat nearby anchor galaxies (for which the binaries may be resolved) in a self-consistent manner to those of more distant galaxies (for which it is not possible to resolve any physical binaries).

For TRGB stars, close binaries might prevent either component from ever getting to the tip because of mass transfer, while wide binaries would have the same effect as self crowding in the general field, that is depleting stars at the tip and moving the pair into the AGB region.  This will blur the tip but not bias its detection \cite{MF_23}.

One can currently only speculate as to the incidence of companions to JAGB/carbon stars. But it is  likely that any surviving (orbitally distant) companions would be fainter (and certainly bluer) than these relatively bright evolved stars. Companions would only compromise the JAGB distance scale if the light contributed by them varied dramatically from galaxy to galaxy, with star formation history differences, or again, with metallicity. However, the small dispersion in the cross-comparison of JAGB distances with TRGB distances to the same galaxies \cite{lee_2023a} puts an upper limit of 0.06~mag on any bias due to the effect of variable contributions from companions to the J-band luminosity function of JAGB stars.

\subsection{Mass Loss}

The masses of Cepheids can be estimated in a number of ways: (1) Using the masses of their main sequence progenitors followed into the Cepheid instability strip, (2) Using stellar pulsation modeling where one of the theoretical input parameters is the mass of the Cepheid and (3) Direct measurement of masses using Cepheids in eclipsing binary systems. For a given period the main sequence masses come in high, the stellar pulsation masses come in low (by 20-30\% \cite{Keller_2008}), and the very rare examples of eclipsing binary stars containing a Cepheid confirm the lower mass estimates (\cite{Piet_2010}). Apparently Cepheids lose mass somewhere between leaving the main sequence and entering the instability strip, at least for the longest-lived second crossing. The (convective) red supergiant phase is the prime suspect, but still unproven. Without knowing the systematics and sensitivities of Cepheid mass loss it can only be speculated as to how much random or systematic noise this one effect is injecting into the observed PL relation as a function of period, age, color and/or metallicity. 

JAGB stars have extended convective envelopes that are unstable to pulsations and in the extreme their redder progeny can develop winds, produce dust and lose mass. A red color cut on the JAGB selection eliminates the reddest stars whose lifetimes are apparently very short, given their small numbers compared to the bluer JAGB color-selected population.

TRGB stars are known to lose mass between the tip and the horizontal branch after the helium flash; but that is after the fact. The TRGB progenitor stars climbing the RGB may or may not be losing mass, but as noted above, the mass of the envelope has little or no effect on the instantaneous or terminal luminosity of these stars.

\subsection{~Boundary Conditions}

For Cepheids, the question of what determines the population of the instability strip remains uncertain and is seldom addressed. Take, for example, the underlying (internal) structure of the Cepheid instability 
strip, and how the triggering of variability (on or off) is externally constrained by the strip's hot/blue and cool/red edges.   What is the physics governing the required depth of the He II ionization layer that turns on the variability as stars enter from the blue? What is the other mechanism that shuts down that same variability as stars exit the instability strip going to the red? And, how are these independent physical mechanisms\footnote{The temperature structure of the envelope in the blue, and the onset of turbulent convection in the red.} controlled by metallicity, surface gravity and intrinsic temperature, say? (Not to mention helium abundance and semi-convection.) If these constraints vary say due to metallicity from galaxy to galaxy, or within galaxies, then the ridge line of the PL relation will be affected in slope and zero point, as will be the entire color/temperature width of the PL relation in which Cepheids might be discovered. See, for example \cite{Somma_2021}.

Boundary conditions controlling the JAGB and/or TRGB phenomena as a whole have already been discussed under their sensitivity to mass, metallicity and evolutionary status.

\begin{figure}
    \centering
\includegraphics[width=7.0cm]{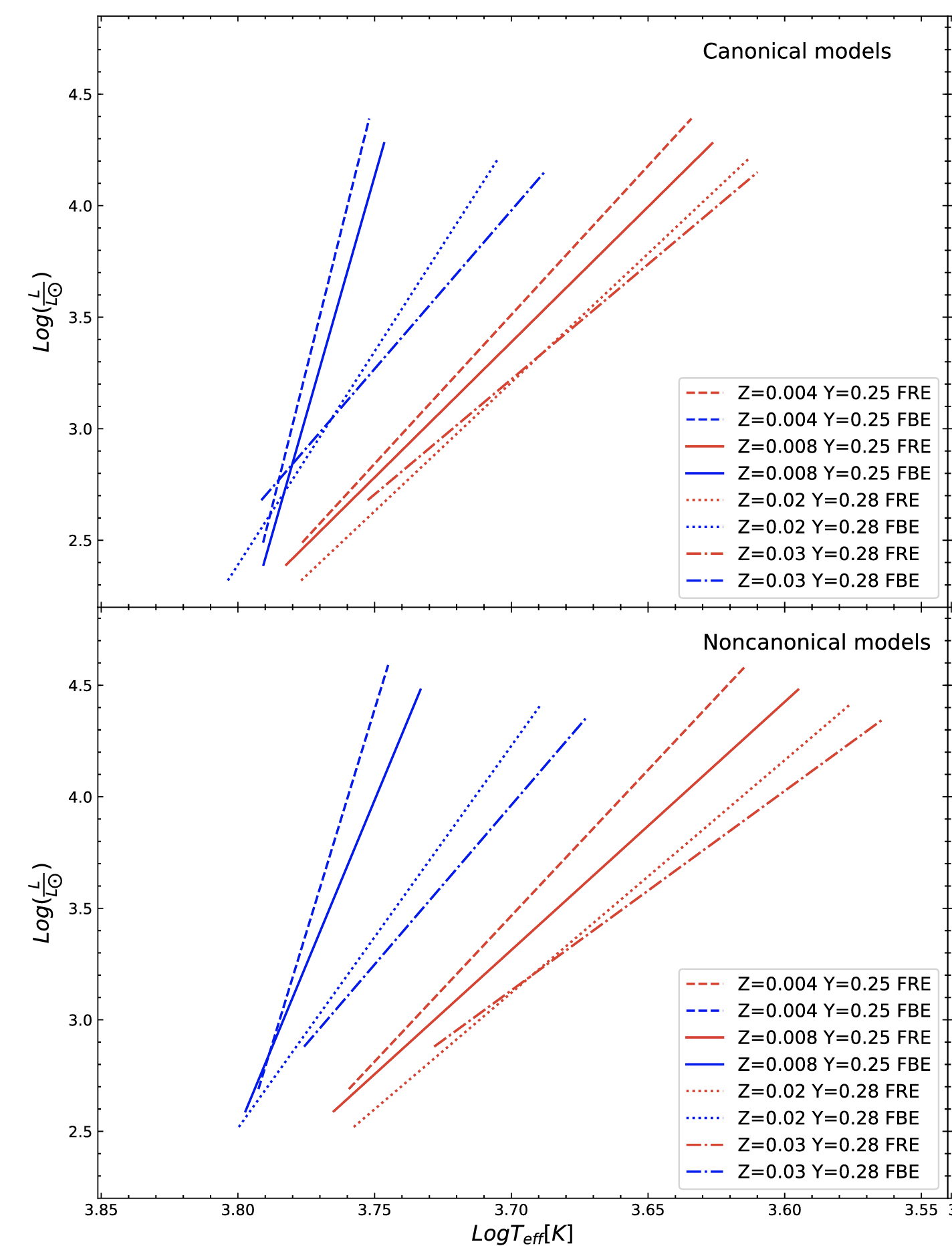}
\caption{Theoretical predictions for the sensitivity of the slopes and intercepts of both the hot (blue) and cool (red) limit of the Cepheid instability strip resulting from changes in the helium abundance and the metallicity (within a single panel) and between adopting canonical and non-canonical mass-luminosity models in the upper and lower panels, respectively. Adapted from \cite{Somma_2021}.}
\label{fig:EDGES}
\end{figure}

\subsection{~Correlated Variance}

The simplest of mathematical considerations requires that for every parameter being solved for there needs to be an independent observation.

For Cepheids we are essentially trying to measure a standardizable intrinsic luminosity in some bandpass, by realizing that the dominant underlying physical parameters controlling the luminosity are temperature, radius and a wavelength-dependent bolometric correction. Interfacing with observations first requires correcting each Cepheid individually, at each wavelength being observed, for total line-of-sight interstellar extinction (in the Milky Way and in the host galaxy). If the atmospheric magnitudes and colors are differentially affected by line blanketing then reddening will be covariant with metallicity. 
Similarly, if metallicity calibrations are based on comparing the distance moduli of galaxies with different mean metallicities then any such calibration will have metallicity and distances being covariant \cite{owens_2022}.

For TRGB stars the main covariance is between luminosity and color (i.e., metallicity) of the tip stars. The correlation has a positive slope (increasing luminosity with increasing color) in the infrared, and a negative slope in the visual and blue. The crossing point of zero slope (i.e., no dependence of magnitude on color) occurs at or slightly redward of the I-band filter near 8100A.

Similarly the J band is where the slope of the JAGB/carbon star absolute magnitude is found to be independent of color (before winds develop and dust forms in the very reddest stars, which are easily excluded by very red color cuts).

\subsection{~Mean Magnitudes}
These final two categories are more on the technical side, and have been left for last because they are well known and their solutions are easily stated and quantified, if not all that easily achieved in practice.

Classical Cepheids can only be uniquely identified by their characteristic asymmetric saw-tooth light curves in the optical, combined with their periods, that can stretch from 2 to more than 100~days. Because their amplitudes monotonically decline with increasing wavelength, Cepheids {\cite{Wisniew_1968} become increasingly harder to detect and characterize in the infrared alone. On the other hand obtaining mean magnitudes good to a given statistical error require fewer observations in the infrared (where random sampling is always closer to the mean in the infrared than are the same phase samples in the optical) than in the blue for example. Be that as it may, several high-precision mean magnitudes obtained for at least two different wavelengths are needed for the Cepheids when correcting their apparent magnitudes for wavelength-dependent interstellar extinction. Four bandpasses were found to be necessary for the \shoes \cite{riess_2022} program of discovery and measurement.

The TRGB magnitude used in this method's distance determination is the magnitude of the discontinuity of the RGB luminosity function corrected for metallicity in all bands except for one, the I band where the slope of the metallicity-color relation changes sign from sloping down to sloping up in moving from the blue to red, effectively crossing zero in the I band around 8100A. Sufficient numbers of TRGB stars are required to fill the luminosity function up to the discontinuity and high-precision data is a benefit (see \cite{MF_23}).
``All that is required" is good areal coverage of the halo and sufficiently long exposures with any given telescope. In principle only a single-epoch exposure in the I band is required to produce the needed apparent luminosity function. In practice two bands are generally required.

The marginalized luminosity function of JAGB stars is optimally undertaken in the J band where the color sensitivity of these color-selected carbon stars is flat with color. While all of these stars are thought to be variables (of one sort or another) the JAGB population is not selected on light curve shapes or any type of variability. Variability is simply another form of random noise that can be averaged over or simply accepted without any systematic penalty.

\subsection{~Optimal Bandpasses}

Measurement of a fiducial TRGB magnitude is optimally undertaken in the I band where the run of tip magnitude with color/metallicity is flattest. Using the tip as a distance indicator at longer or shorter wavelengths requires high-precision colors in order to take out the slope of the tip with color (rectification) without introducing additional noise.

Cepheids require at least two (and optimally three) sets of light curves in order to have time-averaged magnitudes that span a number of different wavelengths. These data points are then needed, in the first instance, to correct for total line-of-sight extinction, and to facilitate the application of metallicity corrections. Two bands in the optical provide good leverage on the reddening, and an additional band as far into the near infrared as possible, is thought to be less influenced by metallicity (but see \cite{Breuval_2022}) and is certainly less impacted by extinction.

The JAGB luminosity function is found to be optimally measured in the J band where the color-selected JAGB candidates are found to show minimal correlation with color.
To date no calibrations have been proposed at any other wavelengths, shorter or longer, given that J-band sensitive instruments are now available on the ground and in space (on both \jwst and \hst}.

\subsection{~Comments}
The various challenges facing the three astrophysical/stellar distance indicators at the foundation of the local distance scale, as discussed individually in some detail above, are summarized in Table 1 below. The three distance indicators are given across the top of the table and each of the topics discussed above are listed from top to bottom. Brief notes describe the challenges facing that particular distance indicator in that particular category, while the details can be found in the corresponding subsections above.

Of the three methods, Cepheids are the most complex and challenging. This is due to their complicated evolutionary status, pulsation properties, mass loss uncertainties, the interplay of interior and atmospheric metallicities across many wavelengths, and finally the vexing issue of crowding that they face in the (dusty) high-surface-brightness inner disks of galaxies where they are confined to be located. Because of their intrinsic simplicity, TRGB stars and JAGB stars are closer to being standard candles when observed in the I and J bands, and when purposefully measured in the halos and in the radially extended disks of their host galaxies, respectively. \jwst observations promise to improve the measurements, as well as to further constrain systematic effects, for all three methods.

\begin{figure}
\centering
\includegraphics[width=13.0cm]{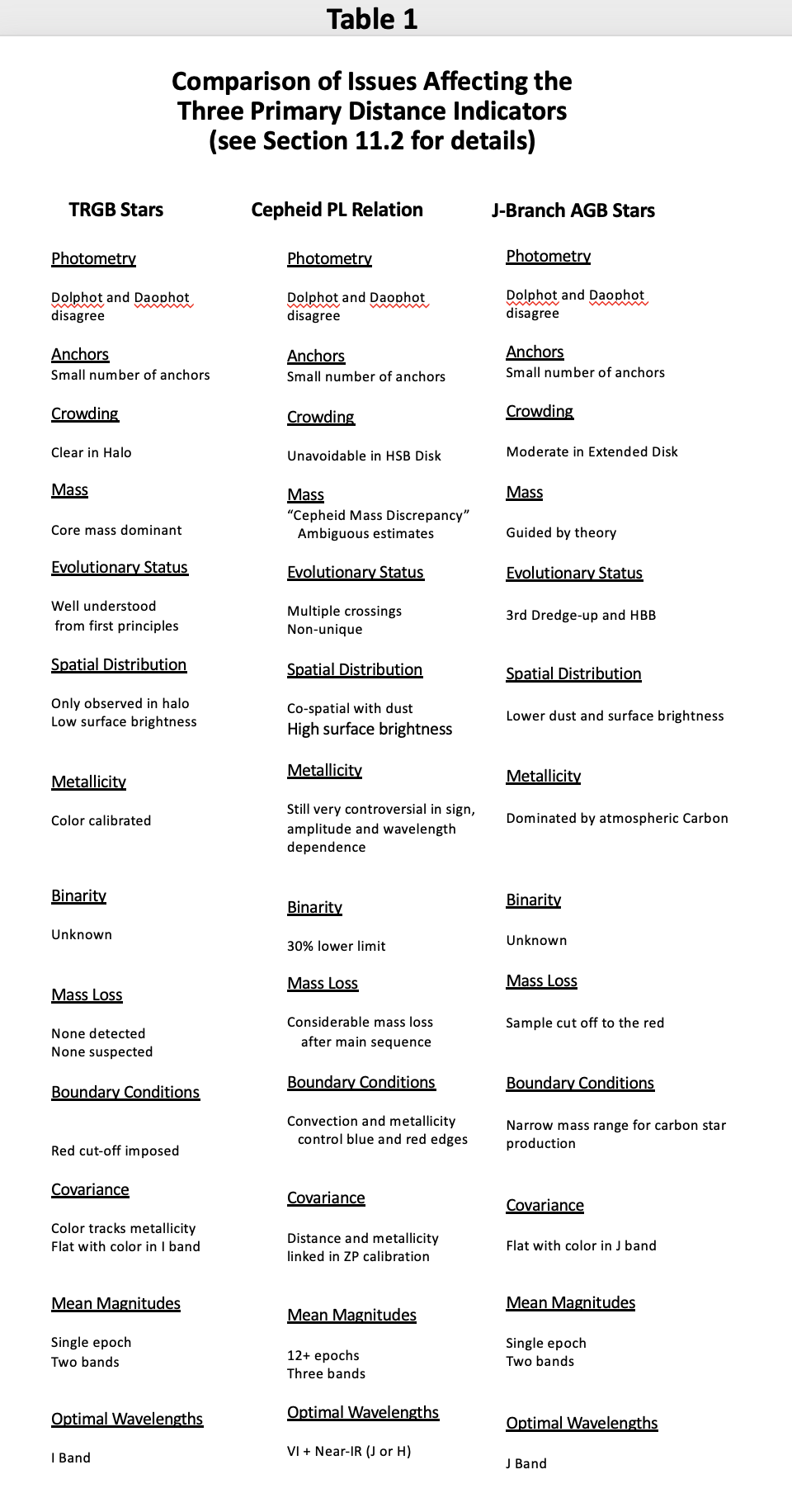}
%\caption{}
%\label{fig:EDGES}
\end{figure}

\acknowledgments

We thank our many students, postdocs, and collaborators over the last 40 years, all of whom contributed to much of the work described here, with particular thanks to Taylor Hoyt, In Sung Jang, Abigail Lee, Andy Monson, Kayla Owens, and Eric Persson. We thank Taylor Hoyt for Figure \ref{fig:Taylor_sn}, and for discussions of the \ho tension. We also thank the University of Chicago and the Carnegie Institution for Science for their support of this research.

This research is based in part on observations made with the NASA/ESA Hubble Space Telescope obtained from the Space Telescope Science Institute, which is operated by the Association of Universities for Research in Astronomy, Inc., under NASA contract NAS 5–26555. 
This work is also based in part on observations made with the NASA/ESA/CSA James Webb Space Telescope. 
The data were obtained from the Mikulski Archive for Space Telescopes at the Space Telescope Science Institute, which is operated by the Association of Universities for Research in Astronomy, Inc., under NASA contract NAS 5-03127.
These observations are associated with HST programs \#12880 and \#14149 and with JWST program \#1995.
Financial support for this work was provided in part by NASA through HST program \#16126 and JWST program \#1995.  

%\paragraph{Note added.} This is also a good position for notes added after the paper has been written.

\clearpage
% Bibliography

%% [A] Recommended: using JHEP.bst file
\bibliographystyle{JHEP}
\bibliography{biblio.bib}

\end{document}